\documentclass[iicol, sn-basic, Numbered]{sn-jnl}

\usepackage{graphicx}%
\usepackage{multirow}%
\usepackage{amsmath,amssymb,amsfonts}%
\usepackage{amsthm}%
\usepackage{mathrsfs}%
\usepackage[title]{appendix}%
\usepackage{xcolor}%
\usepackage{textcomp}%
\usepackage{manyfoot}%
\usepackage{booktabs}%
\usepackage{algorithm}%
\usepackage{algorithmicx}%
\usepackage{algpseudocode}%
\usepackage{listings}%
\usepackage{subcaption}%
\usepackage{adjustbox}%
\usepackage{caption}%

\theoremstyle{thmstyleone}%
\theoremstyle{thmstyletwo}%

\usepackage{xcolor} 
\usepackage{mdframed} 

\theoremstyle{thmstylethree}%

\newtheorem{definitioninner}{Definition}%
\newenvironment{definition}
{\begin{mdframed}[backgroundcolor=gray!20, hidealllines=true] \begin{definitioninner}}
		{\end{definitioninner} \end{mdframed}}
	
\newtheorem{methodology}{Methodology}
\newenvironment{mytheo}
{\begin{mdframed}[backgroundcolor=gray!20, hidealllines=true] \begin{methodology}}
		{\end{methodology} \end{mdframed}}
	
\newtheorem{lemma}{Lemma}
\newenvironment{mylem}
{\begin{mdframed}[backgroundcolor=gray!20, hidealllines=true] \begin{lemma}}
		{\end{lemma} \end{mdframed}}

\begin{document}

\title[Article Title]{Ranking with Ties based on Noisy Performance Data}


\author*[1]{\fnm{Aravind} \sur{Sankaran}}\email{aravind.sankaran@rwth-aachen.de}

\author[2]{\fnm{Lars} \sur{Karlsson}}\email{larsk@cs.umu.se}

\author[2]{\fnm{Paolo} \sur{Bientinesi}}\email{pauldj@cs.umu.se}

\affil*[1]{ \orgname{{RWTH Aachen University}, \city{Aachen}, \country{Germany}}}

\affil[2]{\orgname{Ume\r{a} Universitet}, \country{Sweden}}


\abstract{
  We consider the problem of ranking a set of objects based on their performance when the measurement of said performance is subject to noise. In this scenario, the performance is measured repeatedly, resulting in a range of measurements for each object. 
If the ranges of two objects do not overlap, then we consider one object as ``better'' than the other, and we expect it to receive a higher rank; 
if, however, the ranges overlap, then the objects are \textit{incomparable}, and we wish them to be assigned the same rank.
Unfortunately,
the incomparability relation of ranges is in general not transitive; as a consequence, in general the two requirements cannot be satisfied simultaneously, i.e., it is not possible to guarantee both distinct ranks for objects with separated ranges, and same rank for objects with overlapping ranges. This conflict leads to more than one reasonable way to rank a set of objects.
In this paper,
we explore the ambiguities that arise when ranking with ties,
and define a set of reasonable rankings, which we call {\em partial rankings}. We develop and analyse three different methodologies to compute a partial ranking. 
Finally, we show how 
performance differences among objects can be investigated with the help of partial ranking. 
}




\maketitle

\section{Introduction}
\label{sec3:int}

The problem of ranking a set of objects based on noisy performance measurements appears in various domains, such as High-Performance Computing (HPC) and Business Process Management (BPM). For example, in the field of HPC it is common to compare (i.e., to rank) a set of algorithms based on their execution time, and it is known that execution time measurements typically  exhibit fluctuations due, for instance, to the compute environment~\cite{nikitenko2021influence}. Similarly, in BPM, one might want to rank different workflows within an organization based on non-deterministic throughput times of each workflow. In this paper, we explore and develop methodologies to rank objects based on noisy measurement data while allowing for ties. 

\begin{figure}[h!]
	\centering
	\includegraphics[width=\linewidth]{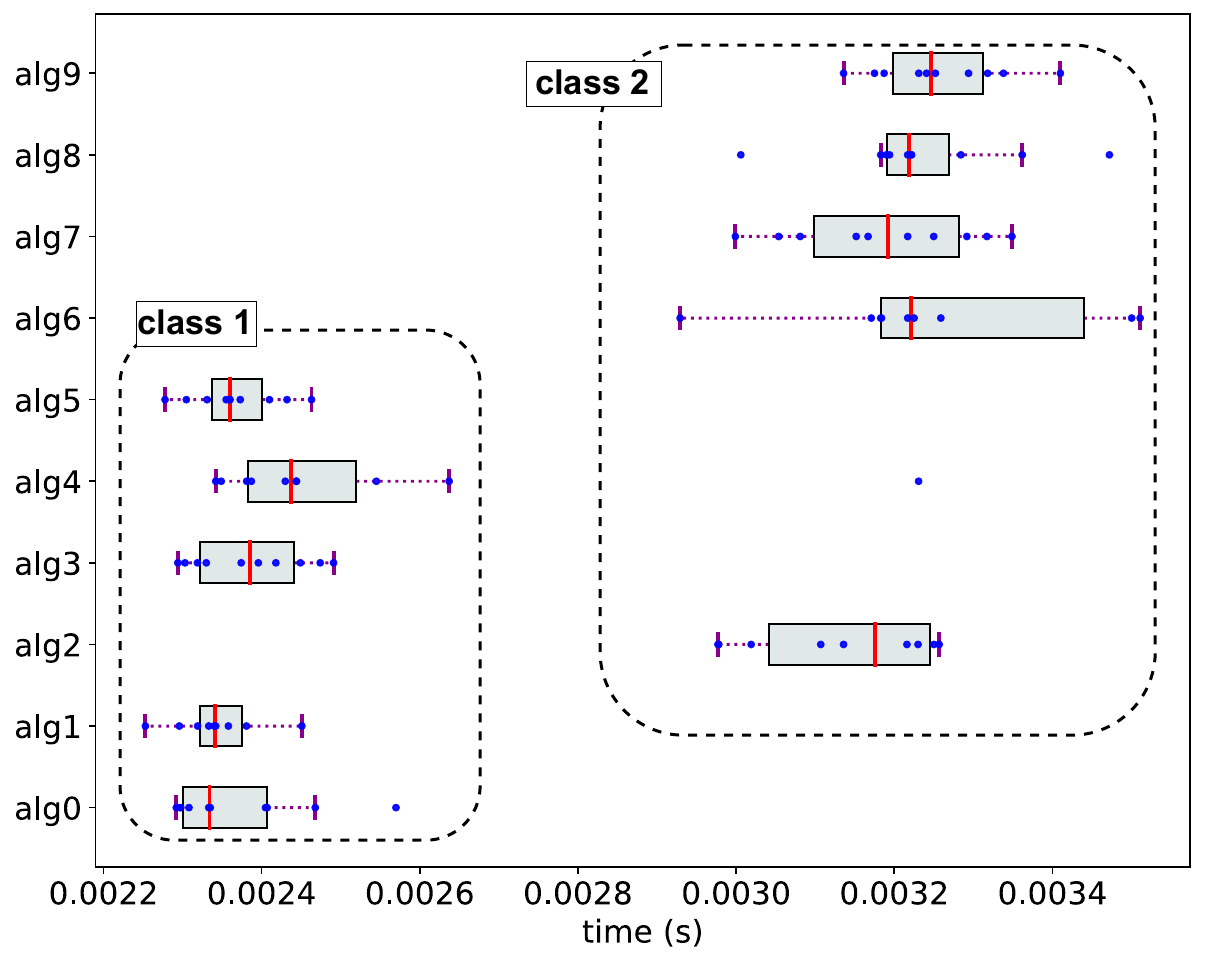}
	\caption{The execution time measurements of ten algorithmic variants to solve the GLS problem:  $(X^{T}M^{-1}X)^{-1}X^{T}M^{-1}\mathbf{y}$ where  $X \in \mathbb{R}^{1000 \times 100}$, $M \in \mathbb{R}^{1000 \times 1000}$ and $\mathbf{y} \in \mathbb{R}^{1000}$. For each variant, the execution times are shown as box plots; red lines indicate the median values; the box indicates the Inter-Quartile Interval. }
	\label{fig3:gls-eg-intro}
\end{figure}

Let us first illustrate the difficulty in ranking with ties. Consider the problem of ranking ten algorithmic variants for the solution of a Generalized Least Squares (GLS) problem based on their execution times, as shown in Fig.~\ref{fig3:gls-eg-intro}
(the variants were automatically generated by the compiler Linnea~\cite{barthels2021linnea} and are mathematically equivalent, i.e., in exact arithmetic they would produce the exact same result). 
The variants were implemented in the Julia language~\cite{bezanson2017julia} and run on a fixed problem size. 
For each variant, the execution time was measured ten times on a Linux-based machine using 12 cores of an Intel-Xeon processor with turbo-boost enabled. The execution time measurements are depicted as box plots, where the box represents the Inter Quartile Interval (IQI), which is the interval between the 25th and the 75th quantile values,  and the red line within the box represents the median value. 

Ranking approaches that rely solely on summary statistics such as minimum or median values often lead to the assignment of a distinct rank to each variant, thus overlooking the possibility of ties (except when the statistics are exactly identical). By contrast, a ranking method that accounts for ties can group variants into \textit{performance classes}, where each class includes variants with the same rank. 
Consider the two performance classes marked in Fig.~\ref{fig3:gls-eg-intro} based on the better-than relation $<_{eg}$, which takes into account ties among the variants as follows: a variant $\mathbf{alg}_i$ is considered to be better than another variant $\mathbf{alg}_j$ (i.e., $\mathbf{alg}_i <_{eg} \mathbf{alg}_j$) if and only if the IQI of $\mathbf{alg}_i$ lies entirely to the left of the IQI of $\mathbf{alg}_j$.  If the IQI of $\mathbf{alg}_i$ and $\mathbf{alg}_j$ overlap with one another, then  $\mathbf{alg}_i$ and $\mathbf{alg}_j$ are considered \textit{incomparable} (i.e., $\mathbf{alg}_i \sim \mathbf{alg}_j$).

In typical ranking problems, it is assumed that the better-than relation follows a \textit{strict weak ordering} (i.e., not allowing for non-transitive ties). However, non-transitive ties can be expected in a noisy measurement data; in our example, $\mathbf{alg}_{1} \sim \mathbf{alg}_{3}$ and $\mathbf{alg}_{3} \sim \mathbf{alg}_{4}$, but $\mathbf{alg}_{1} <_{eg} \mathbf{alg}_{4}$. Therefore, the following ambiguity arises in the ranking process: Should $\mathbf{alg}_{4}$ and $\mathbf{alg}_{1}$ be placed in different ranks because their intervals (IQIs) are separated, or should both of them be assigned the same rank because their IQIs mutually overlap with that of $\mathbf{alg}_{3}$? 

\begin{figure}[h!]
	\centering
	\includegraphics[width=\linewidth]{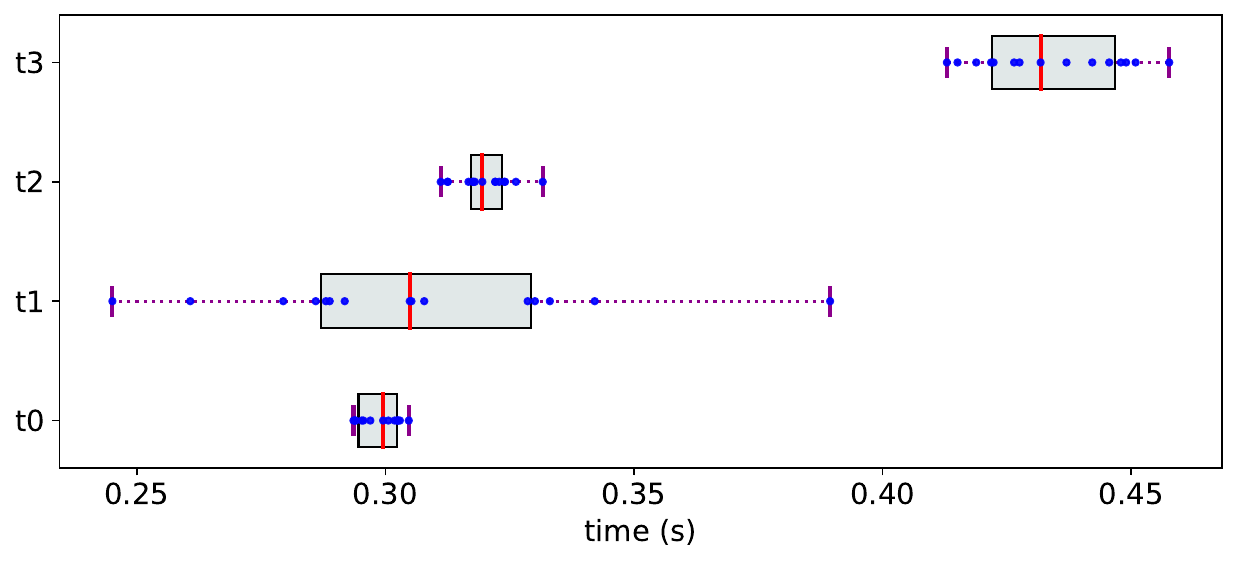}
	\caption{Sets of measurements ($\mathcal{M}_0$).}
	\label{fig3:intro-eg}
\end{figure}

In general, it is possible to have several reasonable rankings for given sets of measurement data and a better-than relation. To illustrate how different rankings can be obtained through various arguments, let us utilize the following simulated example. Consider the set of variants  $\mathcal{M} = \{\mathbf{t}_0, \mathbf{t}_1, \mathbf{t}_2, \mathbf{t}_3\}$ with each $\mathbf{t}_i \in \mathcal{M}$ indicated as a set of measurement values $\mathbf{t}_i\in \mathbb{R}^{M}$ sampled from normal distribution $\mathcal{N}(\mu, \sigma)$ that simulates the cost of the variant. Here, $\mu$ represents the mean and $\sigma$ represents the standard deviation of the normal distribution. Let us assume that the variants are characterized by the following distributions:
\begin{itemize}
	\setlength{\itemsep}{0pt} 
	\item $\mathbf{t}_0$: Sampled from $\mathcal{N}(0.30,0.005)$.
	\item $\mathbf{t}_1$: Sampled from $\mathcal{N}(0.31,0.030)$.
	\item $\mathbf{t}_2$: Sampled from $\mathcal{N}(0.32,0.005)$.
	\item $\mathbf{t}_3$: Sampled from $\mathcal{N}(0.43, 0.01)$.
\end{itemize}
For each variant, $M=15$ values are sampled and the resulting set of values are depicted in Fig.~\ref{fig3:intro-eg} as box plots.

For the sets of measurements ($\mathcal{M}_0$) in Fig.~\ref{fig3:intro-eg}, the relation $<_{eg}$ among the variants is shown as a (transitively reduced) directed graph in Fig.~\ref{fig:intro-dfgl}; a directed path from $\mathbf{t}_i$ to $\mathbf{t}_j$ indicates that $\mathbf{t}_i <_{eg} \mathbf{t}_j$. 
An example of the ambiguity for this case could be that the three rankings with ties shown in Fig.~\ref{fig:intro-rank} are reasonable by the following justifications:

\begin{figure*}[h!]
	\centering
	
	\begin{subfigure}[b]{0.48\textwidth}
		\centering
		\includegraphics[width=0.3\linewidth]{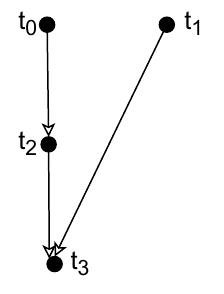}
		\caption{$<_{eg}$ on $\mathcal{M}_0$ shown as a transitively reduced directed graph.}
		\label{fig:intro-dfgl}
	\end{subfigure}
	\hfill
	\begin{subfigure}[b]{0.48\textwidth}
		\includegraphics[width=0.8\linewidth]{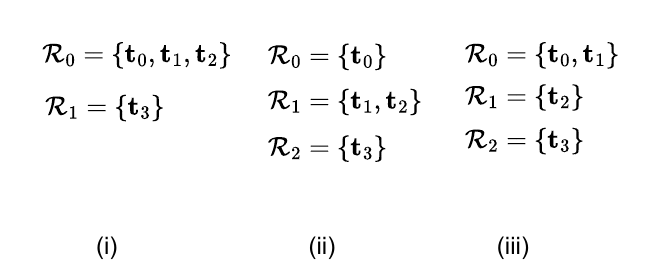}
		\caption{Reasonable rankings deduced from Fig.~\ref{fig:intro-dfgl}. $\mathbf{t}_i \in \mathcal{R}_k$ implies $\mathbf{t}_i$ is assigned the rank $k$.}
		\label{fig:intro-rank}
	\end{subfigure}
	\caption{When distributions overlap, the partial ordering of the algorithms admits many reasonable rankings.}
	\label{fig:intro}
\end{figure*}

\begin{itemize}
	\setlength\itemsep{0.1em}
	\item 
	One could argue that since $\mathbf{t}_0 \sim \mathbf{t}_1$ and $\mathbf{t}_1 \sim \mathbf{t}_2$, all $\mathbf{t}_0$, $\mathbf{t}_1$ and $\mathbf{t}_2$  should be given the same rank and considered as the best variants   (Fig~\ref{fig:intro-rank}(i)).
	
	\item Alternatively, one could argue that $\mathbf{t}_0$ should be ranked higher than $\mathbf{t}_2$ because $\mathbf{t}_0 <_{eg} \mathbf{t}_2$, and
	$\mathbf{t}_1$ should be ranked as low as $\mathbf{t}_2$ because the IQI of $\mathbf{t}_1$ overlaps with that of $\mathbf{t}_2$ (Fig~\ref{fig:intro-rank}(ii)). Then, only $\mathbf{t}_0$ should be considered as the best variant.
	
	\item Alternatively, since  $\mathbf{t}_0 \sim \mathbf{t}_1$ and $\mathbf{t}_0 <_{eg} \mathbf{t}_2$, one could argue that both $\mathbf{t}_0$ and $\mathbf{t}_1$ should be considered as the best variants, and $\mathbf{t}_2$ should be placed in a lower rank than $\mathbf{t}_0$  (Fig~\ref{fig:intro-rank}(iii)).
	
\end{itemize}

Thus, as soon as one considers a better-than relation that induces non-transitive ties, multiple reasonable rankings emerge. In this paper, we first define what constitutes the set of reasonable rankings in general, and then develop methodologies to compute such rankings. Instead of a strict weak ordering, we approach the ranking problem based on better-than relations that follow \textit{strict partial ordering}, and therefore call the resulting rankings as \textit{partial rankings}. We then apply the methodology to the identification of the root cause of performance differences among the objects.
The contributions of this work are the following:
\begin{enumerate}
	\setlength{\itemsep}{0pt} 
	
	\item  For given sets of measurements and a better-than relation that defines how two sets of  measurements should be compared, we define partial ranking to identify a set of reasonable rankings.
	
	\item We develop three different methodologies to compute some partial rankings for a given sets of measurements according to a given better-than relation.
	
	\item We show how our partial ranking methodologies can be used to discover the underlying causes of performance differences between the objects.
	
\end{enumerate}

\paragraph{\textit{Organization: }} In Sec.~\ref{sec3:def}, we explain and formally define partial ranking. In Sec.~\ref{sec3:rel}, we discuss the related works. In Sec.~\ref{sec3:met}, we present methodologies for partial ranking of sets of measurements, and in Sec.~\ref{sec3:handlingq}, we extend our methodologies to handle multiple better-than relations. In Sec.~\ref{sec3:exp}, we apply the partial ranking methodology in two practical scenarios, and finally, in Sec.~\ref{sec3:con}, we draw conclusions.

\section{Partial Ranking}
\label{sec3:def}

Let $\mathcal{M} = \{\mathbf{t}_0, \dots, \mathbf{t}_{N-1}\}$ be a set of $N$ objects. Let $<_{\mathbf{P}} $ be a \textit{strict partial order} on $\mathcal{M}$ that models a better-than relation
between a pair of objects  $\mathbf{t}_i, \mathbf{t}_j \in \mathcal{M}$, i.e.,  $\mathbf{t}_i <_{\mathbf{P}}  \mathbf{t}_j$  means that $\mathbf{t}_i$ is somehow \textit{better than} (e.g., faster than) $\mathbf{t}_j$. If  neither $\mathbf{t}_i <_{\mathbf{P}}  \mathbf{t}_j$ nor  $\mathbf{t}_j <_{\mathbf{P}}  \mathbf{t}_i$, then $\mathbf{t}_i$ and $\mathbf{t}_j$ are \textit{incomparable}, which we denote by  $\mathbf{t}_i \sim \mathbf{t}_j$.
For each $\mathbf{t}_i \in \mathcal{M}$, we aim to assign a rank in the form of a non-negative integer with 0 representing the highest (best) rank that is consistent with $<_{\mathbf{P}} $. 
We first make a distinction between the cases where (1) we want the ranking to be unique and without ties, (2) we want the ranking to be unique but with ties, (3) the ranking cannot be unique but allows ties. 

\begin{figure*}[h!]
	\centering
	\begin{subfigure}[b]{0.59\textwidth}
		\centering
		\includegraphics[width=1\linewidth]{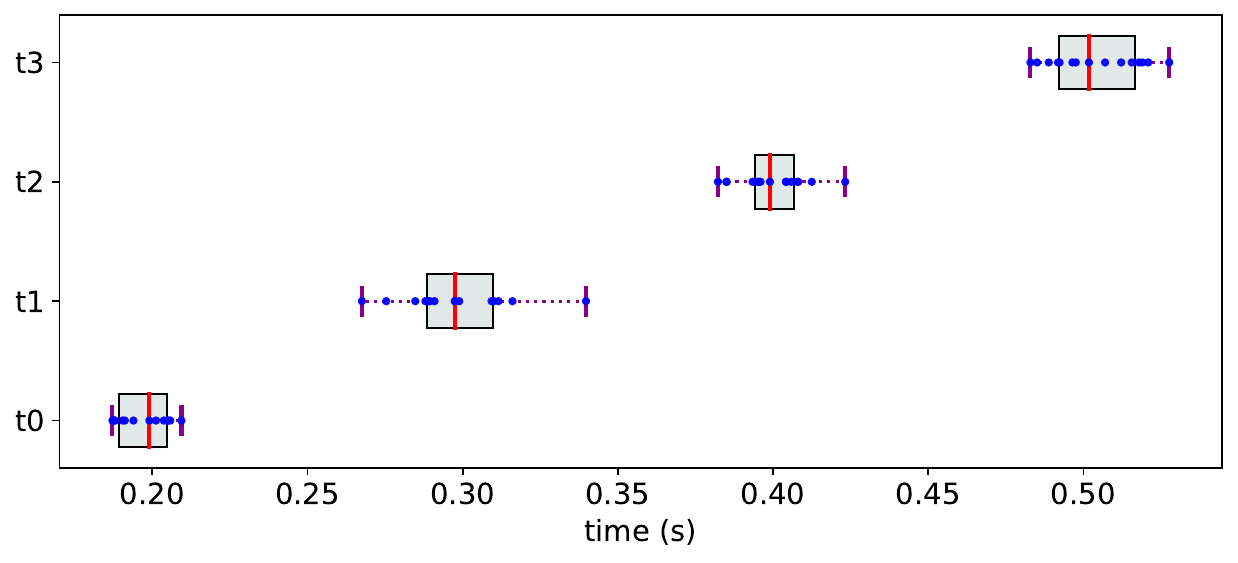}
		\caption{Sets of measurements $\mathcal{M}_1$.}
		\label{fig:total-eg}
	\end{subfigure}
	\begin{subfigure}[b]{0.39\textwidth}
		\centering
		\includegraphics[width=0.6\linewidth]{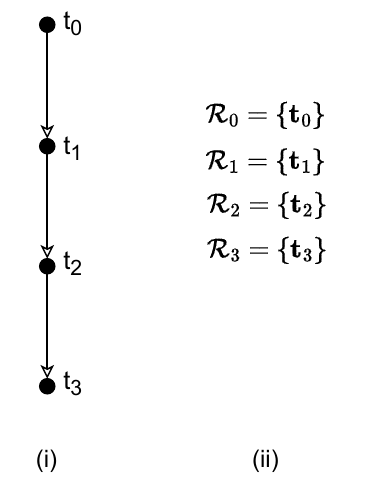}
		\caption{$<_{eg}$ on $\mathcal{M}_1$ and the ranking (Linear order).}
		\label{fig:total}
	\end{subfigure}
	\par\bigskip
	\begin{subfigure}[b]{0.59\textwidth}
		\includegraphics[width=1\linewidth]{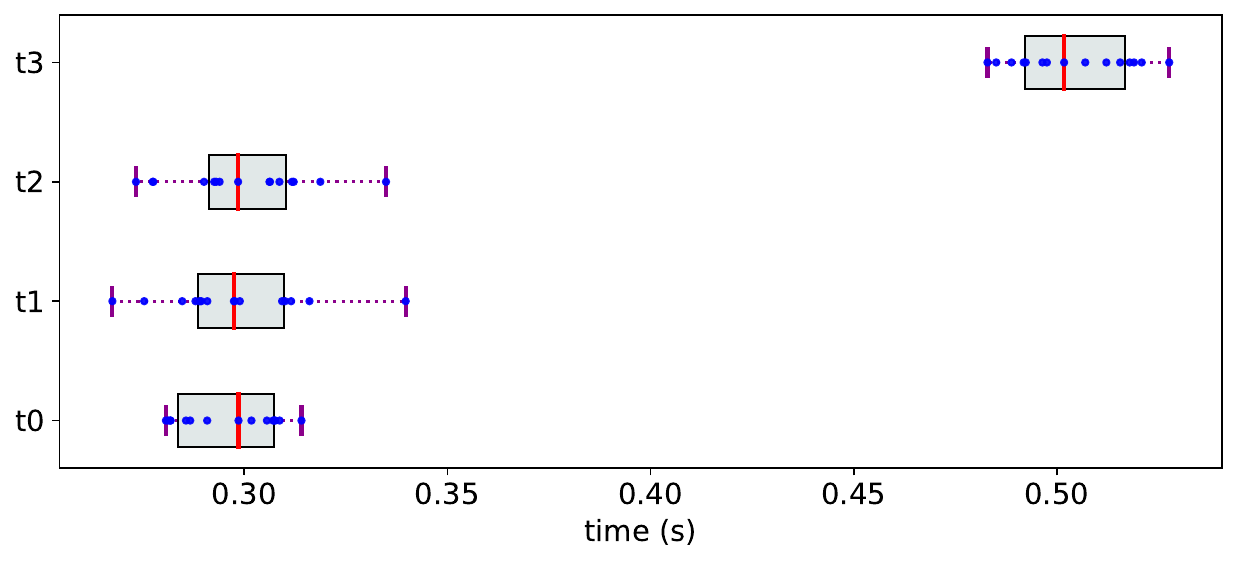}
		\caption{Sets of measurements $\mathcal{M}_2$.}
		\label{fig:weak-eg}
	\end{subfigure}
	\begin{subfigure}[b]{0.39\textwidth}
		\centering
		\includegraphics[width=0.85\linewidth]{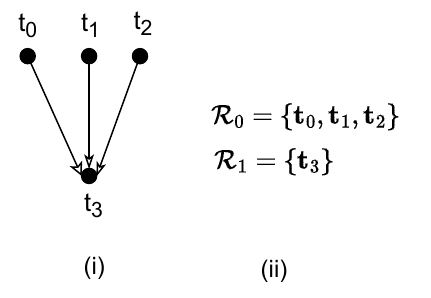}
		\caption{$<_{eg}$ on $\mathcal{M}_2$ and the ranking (Weak order).}
		\label{fig:weak}
	\end{subfigure}
	
	\caption{When $<_{\mathbf{P}} $ imposes either a linear order or a weak order on $\mathcal{M}$, then we want the ranking to be unique. Moreover, when the order is linear, then we want the ranking to have no ties.}
	\label{fig:pr-def}
\end{figure*}
\begin{enumerate}
	\setlength\itemsep{1em}
	\item \textbf{Linear order}  : If $\forall \mathbf{t}_i, \mathbf{t}_j \in \mathcal{M}$, either $\mathbf{t}_i <_{\mathbf{P}}  \mathbf{t}_j$ or  $\mathbf{t}_j <_{\mathbf{P}}  \mathbf{t}_i$, then $\mathcal{M}$ follows a \textit{linear order} for $<_{\mathbf{P}} $. In other words, $\nexists \mathbf{t}_i, \mathbf{t}_j \in \mathcal{M}$ such that $\mathbf{t}_i \sim \mathbf{t}_j$. In this case, we want the ranking to be unique and without ties.
	
	\noindent \textit{Example:} Consider again the four sets of measurements $\mathcal{M}_0$ shown in  Fig.~\ref{fig3:intro-eg} and the better-than relation $<_{med}$ where $\mathbf{t}_i <_{med} \mathbf{t}_j$ if and only if the median of $\mathbf{t}_i$ is smaller than the median of $\mathbf{t}_j$. According to $<_{med}$, the resulting ranking is the ordered set partition  ---$\mathcal{R}_0=\{\mathbf{t}_0\}$, $\mathcal{R}_1 = \{\mathbf{t}_1\}$, $\mathcal{R}_2 = \{\mathbf{t}_2\}, \mathcal{R}_3 = \{\mathbf{t}_3\}$--- of $\mathcal{M}_0$.
	Thus, $\mathcal{M}_0$ follows linear order for $<_{med}$.
	Consider another example $\mathcal{M}_1$ shown in Fig.~\ref{fig:total-eg} and the better-than relation $<_{eg}$. 
	Here, $<_{eg}$ imposes linear order on $\mathcal{M}_{1}$, and the resulting ranking is shown in Fig.~\ref{fig:total}(ii).

	\item \textbf{Weak order}: If $\forall \mathbf{t}_i, \mathbf{t}_j, \mathbf{t}_k \in \mathcal{M}$ such that $\mathbf{t}_i \sim \mathbf{t}_j$ and $\mathbf{t}_j \sim \mathbf{t}_k$, it holds $\mathbf{t}_i \sim \mathbf{t}_k$ (in  other words, the incomparability relation is transitive),  then $\mathcal{M}$ follows a \textit{weak order} for $<_{\mathbf{P}} $.   In this case,  we want the ranking to be unique, but this time with ties among all objects in the same equivalence class induced by the incomparability relation. 
	
	\noindent \textit{Example:} Consider the four sets of measurements $\mathcal{M}_2$ shown in Fig.~\ref{fig:weak-eg} ordered by $<_{eg}$. The corresponding relations is shown in Fig.~\ref{fig:weak}(i). The resulting ranking is illustrated in Fig.~\ref{fig:weak}(ii). Here,  $\mathbf{t}_0, \mathbf{t}_1, \mathbf{t}_2 \in \mathcal{R}_0$ are all pairwise incomparable and all are assigned the same rank.
	
	\item \textbf{Neither linear nor weak}: If $\exists \mathbf{t}_i, \mathbf{t}_j, \mathbf{t}_k \in \mathcal{M}$ such that $\mathbf{t}_i \sim \mathbf{t}_j$, $\mathbf{t}_j \sim \mathbf{t}_k$ and $\mathbf{t}_i <_{\mathbf{P}} \mathbf{t}_k$ (in  other words, the incomparability relation is \textit{not} transitive), then $\mathcal{M}$ is neither linear nor weak for $<_{\mathbf{P}} $. In this case, the ranking will not be unique, for the reasons outlined in the introduction.

	\noindent \textit{Example}: The sets of measurements $\mathcal{M}_0$ in Fig.~\ref{fig3:intro-eg} follows neither linear nor weak order for $<_{eg}$. For this case, we expect one of the three rankings shown in Fig.~\ref{fig:intro-rank}. 
	
\end{enumerate}

We now formulate a definition for ranking with ties that results in unique rankings when $<_{\mathbf{P}} $  imposes a linear or weak order on $\mathcal{M}$, and in more general cases allows only those rankings that are reasonable, i.e., they satisfy the following properties.

\begin{figure}[h!]
	\centering
	\includegraphics[width=0.4\linewidth]{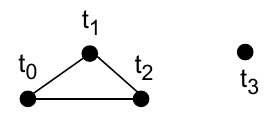}
	\caption{Undirected graph of the objects in $\mathcal{M}_2$ associated by $\sim$ according to $<_{eg}$.}
	\label{fig3:undirected}
\end{figure}

\begin{definition}
	\label{th:ranking}
	Given a set of objects $\mathcal{M}$ and a strict partial order relation $<_{\mathbf{P}} $ on $\mathcal{M}$, we define a \textbf{\textit{partial ranking}} as an ordered set partition $\mathcal{R}_0, \dots, \mathcal{R}_{K-1}$ of $\mathcal{M}$ with the following properties: 
	
	\begin{enumerate}
		\setlength\itemsep{0.5em}

		\item   If $\mathbf{t}_i$ is ranked higher than $\mathbf{t}_j$, then either $\mathbf{t}_i <_{\mathbf{P}}  \mathbf{t}_j$ or $\mathbf{t}_i \sim \mathbf{t}_j$.
		
		\item For all pairs of consecutive ranks $(\mathcal{R}_a, \mathcal{R}_{a+1})$, there exists a $\mathbf{t}_i \in \mathcal{R}_a$ and $\mathbf{t}_j \in \mathcal{R}_{a+1}$ such that $\mathbf{t}_i <_{\mathbf{P}}  \mathbf{t}_j$.

		\item For all pairs $\mathbf{t}_i$ and $\mathbf{t}_j$ with the same rank, $\mathbf{t}_i$ and $\mathbf{t}_j$ must be connected in the undirected graph associated with the incomparability relation.

	\end{enumerate}
\end{definition}
\bigskip
\noindent The three properties serve the following purposes:
\begin{itemize}
	\item   Property~1 ensures that the ranking is consistent with the strict partial order.
	This property captures the essence of what we consider a reasonable ranking, since it prevents an object that is better than another from being ranked lower than the other.
	\item  Property~2 prohibits splitting up mutually incomparable objects into separate ranks. This property is essential to make the ranking unique for weak orders. For example, without this property, when $\mathcal{M}_2$ is ranked according to $<_{eg}$, the ranking $\mathcal{R}_0 = \{\mathbf{t}_0\}, \ \mathcal{R}_1 = \{\mathbf{t}_1, \mathbf{t}_2\}, \ \mathcal{R}_2 = \{\mathbf{t}_3\}$ would have been considered valid as it satisfies both Property 1 and Property 3. However, this ranking is not valid according to Property~2 because there does not exist a $ \mathbf{t}_i \in \mathcal{R}_0$ and $\mathbf{t}_j \in \mathcal{R}_1$ such that $\mathbf{t}_i <_{\mathbf{P}} \mathbf{t}_j$.

	\item  Property~3 ensures that the objects with the same rank cannot be trivially partitioned into separate ranks. This property is essential for the uniqueness for both linear and weak orders, since without this property it would always be possible to assign all objects the same rank.  For example, without this property, for $\mathcal{M}_2$ and $<_{eg}$, the ranking $\mathcal{R}_0 = \{\mathbf{t}_0, \mathbf{t}_1, \mathbf{t}_2, \mathbf{t}_3\}$ would be considered valid as it satisfies both Property 1 and Property~2. However, according to Property~3, $\mathbf{t}_3$ cannot exist in the same rank as the other objects because $\mathbf{t}_3$ is disconnected in the undirected graph shown in Fig.~\ref{fig3:undirected}.
\end{itemize}
In summary, given $\mathcal{M}$ and $<_{\mathbf{P}}$, a ranking that satisfies the properties listed above is called a partial ranking. Now, the problem is to find a procedure that takes $\mathcal{M}$ as input, makes comparisons according to $<_{\mathbf{P}}$, and returns a partial ranking as output.

\section{Related Works}
\label{sec3:rel}


The problem of ranking with ties is generally seen as the partial ordering of a set of objects, in which the objects are related to one another based on a certain notion of importance~\cite{bruggemann2011ranking, janicki2008ranking}.
Although rankings based on partial orders have been widely discussed, we still notice a lack of clarity on what constitutes a reasonable ranking, especially when the partial order is neither linear nor weak. The term partial ranking has been previously used is some works (e.g.,~\cite{fagin2006comparing,ailon2010aggregation, pavan2004new}) to denote a ranking deduced from partial orders, but without much discussion of the properties of those rankings. In this work, we elaborate on partial ranking and explicitly list out its properties.

When it comes to methodologies for ranking based on partial orders, Pavan et al.~\cite{pavan2004new} explored the use of Hasse diagram technique to infer the ranks; i.e., a graph similar to the one shown in Fig~\ref{fig:intro-dfgl} is constructed, and the depth of a node in the graph is interpreted as the rank of that node. However, the ranking problem is addressed in the context of multi-attribute decision making, and the data is not considered as sets of measurements.  In~\cite{sankaran2022test, sankaran2021performance}, the Bubble-sort algorithm is used to develop a ranking methodology that allows for ties based on incomparable pairs of sets of measurements. However, the Bubble-sort based methodology is suited for a strict weak ordering and does not always compute a partial ranking. 

There is a large body of literature on the area of ranking, particularly tailored for applications in information retrieval; for example,  among the database community, \cite{ilyas2008survey} surveys the top-k ranking techniques for information retrieval and ~\cite{li2010ranking} discusses ranking approaches based on datasets that exhibit uncertainty. However, these works do not explicitly address the inherent ambiguities in rankings caused by noisy measurement data. Statistical ranking approaches such as~\cite{szekli1995stochastic} often require assumptions to be made regarding the underlying distribution of the measurement values, and the number of measurement values (or the sample size) plays an important role in determining whether or not the resulting rankings are meaningful.
However, in many situations, measurements do not follow  standard distribution; for instance, the measurements of execution times in a compute system  are generally multi-modal (i.e., the measurement values occur in stratified clusters, mainly due to the processor operating at multiple frequency levels~\cite{charles2009evaluation}). As a result, it has been noted that the application of  textbook statistical approaches in summarizing such execution time measurements is not straightforward~\cite{chen2015statistical,hoefler2015scientific}. In this work, we consider the fact that the measurements could be noisy, but we do not make any assumption about their underlying distributions. Moreover, we focus on approaches that make use of the data---no matter how insufficient they are---as best as possible. 

In order to address the challenges posed by small sample sizes and the variability in measurement values, the spread of measurements is sometimes summarized using interval numbers~\cite{zhang2016ranking} or grey numbers~\cite{liu2014ranking}. While ranking approaches based on interval numbers have been applied in previous studies~\cite{liu2014ranking, ye2016method}, they have again focused mainly on the context of multi-attribute decision-making. In this paper,  we use the concept of interval numbers to compare two sets of measurements, but our focus is specifically on the ranking problem rather than decision-making. We clarify the ambiguities in the ranking by adopting our partial ranking definition. 

\section{Methodologies For Partial Ranking}
\label{sec3:met}

In this section, we develop methodologies to compute a partial ranking for a given $\mathcal{M}$ and $<_{\mathbf{P}}$. Let each $\mathbf{t}_i \in \mathcal{M}$ consist of $M_i$ measurement values (i.e.,  $\mathbf{t}_i \in \mathbb{R}^{M_i}$), and we assume that all $M_i \ge 1$.

\subsection{Methodology 1: For an arbitrary number of ranks}
\label{sec3:m2}

For a given $\mathcal{M}$ and $<_{\mathbf{P}}$, let $G$ be a directed graph such that $\mathbf{t}_i \in \mathcal{M}$ are the nodes, and a directed edge from $\mathbf{t}_i$ to $\mathbf{t}_j$ exists if and only if $\mathbf{t}_i <_{\mathbf{P}} \mathbf{t}_j$. A partial ranking can be computed from $G$ as follows:
\bigskip
\begin{mytheo}
	\label{th:problem2}
	Given the graph $G$ constructed from the partial order $(\mathcal{M}, <_{\mathbf{P}})$, set the rank of $\mathbf{t}_i \in \mathcal{M}$ equal to the length of the longest directed path in $G$ that ends at $\mathbf{t}_i$.
\end{mytheo}
\bigskip

Note that $G$ does not contain any cycles as $<_{\mathbf{p}}$ is transitive,
and therefore, $G$ is a Directed Acyclic Graph. As a consequence,  the length of the longest directed path in $G$ that ends at $\mathbf{t}_i$ (or the \textit{depth} of $\mathbf{t}_i$ in $G$), denoted as $d(\mathbf{t}_i)$,  can be computed using the following recursive formula:
\begin{equation}
	d(\mathbf{t}_i) = 
	\begin{cases}
		\underset{\mathbf{t}_j \in \bullet \mathbf{t}_i}{\max }  \ d(\mathbf{t}_j) + 1, & \text{if} \ |\bullet \mathbf{t}_i| > 0 \\
		0, & \text{if}  \ |\bullet \mathbf{t}_i| = 0
	\end{cases}
\end{equation}
where $\bullet \mathbf{t}_i$ is the set of incoming edges to $\mathbf{t}_i$. Then, the ordered set partition of $\mathcal{M}$ is:
\begin{equation}
	\label{eq:rank-hesse}
	\mathcal{R}_k = \{\mathbf{t}_i \in \mathcal{M}\ | \  d(\mathbf{t}_i) = k \} \qquad \forall k \in \{0, \dots, K-1\}
\end{equation}
where  $K-1$ is the length of the longest directed path in $G$. 

\begin{figure}[h!]
	\centering
	\includegraphics[width=\linewidth]{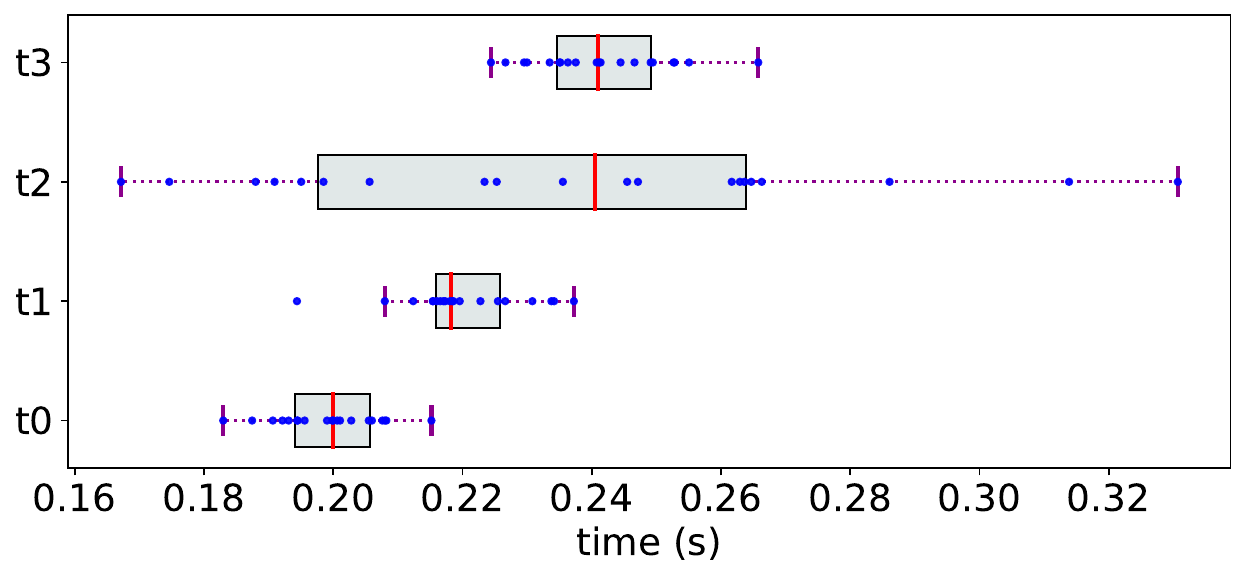}
	\caption{Sets of measurements $\mathcal{M}_3$. The box indicates the IQI.}
	\label{fig3:met-eg}
\end{figure}

\begin{figure}[h!]
	\centering
	\begin{subfigure}{0.45\textwidth}
		\centering
		\includegraphics[width=0.5\textwidth]{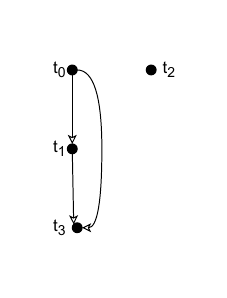}
		\caption{The graph $G$ associated with ($\mathcal{M}_3, <_{eg}$) }
		\label{fig3:met-dfg}
	\end{subfigure}
	\hfill
	\begin{subfigure}{0.42\textwidth}
		\centering
		\includegraphics[width=0.5\textwidth]{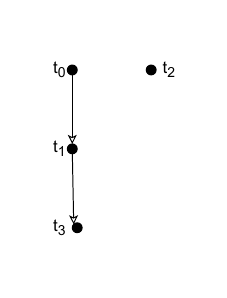}
		\caption{Transitive reduction of $G$.}
		\label{fig3:met-dfg-tr}
	\end{subfigure}
	\caption{Directed graphs from the partial order ($\mathcal{M}_3, <_{eg}$)}
	\label{fig3:met-eg-g}
\end{figure}

For example, consider the sets of measurements $\mathcal{M}_3$ in Fig.~\ref{fig3:met-eg} and the relation $<_{eg}$. $\forall \mathbf{t}_i, \mathbf{t}_j \in \mathcal{M}_3$, we perform pair-wise comparison according to $<_{eg}$ and construct $G$ (shown in Fig.~\ref{fig3:met-dfg}).  Note that performing transitivity reduction on a directed graph removes redundant edges without altering the depth of the nodes.  Consequently, it does not impact the ranking of the nodes. The transitive reduction of $G$ is shown in Fig.~\ref{fig3:met-dfg-tr}. In both Fig.~\ref{fig3:met-dfg} and Fig.~\ref{fig3:met-dfg-tr},
the depth of $\mathbf{t}_0$ and $\mathbf{t}_2$ is 0, the depth of $\mathbf{t}_1$ is 1 and the depth of $\mathbf{t}_3$ is the maximum of $\{2,1\}$ which is 2. Hence, the ordered set partition is $\mathcal{R}_0 = \{\mathbf{t}_0, \mathbf{t}_2\}$, $\mathcal{R}_1 = \{ \mathbf{t}_1\}$, $\mathcal{R}_2 = \{\mathbf{t}_3\}$.

\begin{mylem}
	Rankings produced by Methodology~\ref{th:problem2} are partial rankings (Def.~\ref{th:ranking}). 
\end{mylem}
\begin{proof}
	We need to prove that the ranking follows the three properties in Definition~\ref{th:ranking}.
	\begin{enumerate}
		\item Proof for Property 1:
		\begin{itemize}
			\item  $\forall \mathbf{t}_i \in \mathcal{R}_a$ and  $\forall \mathbf{t}_j \in \mathcal{R}_b$, $a<b \implies  d(\mathbf{t}_i) < d(\mathbf{t}_j) $
			\item $d(\mathbf{t}_i) < d(\mathbf{t}_j) $ $\implies$ $\mathbf{t}_i <_{\mathbf{P}} \mathbf{t}_j$ or $\mathbf{t}_i \sim \mathbf{t}_j$.
		\end{itemize}
		
		\item Proof for Property 2: $\forall \mathcal{R}_a, \mathcal{R}_{a+1} \implies$ there exists a directed edge from some $\mathbf{t}_i \in \mathcal{R}_a$ to some   $\mathbf{t}_j \in \mathcal{R}_{a+1}$, which means $\exists \mathbf{t}_i <_{\mathbf{P}} \mathbf{t}_j$.
		\item Proof for Property 3: Observe that in a directed acyclic graph, two nodes at the same depth cannot be connected by an edge, because if they were, then one of the nodes would be at a depth greater than the other node. Thus, $\forall \mathcal{R}_a$, $ \mathbf{t}_i, \mathbf{t}_j \in \mathcal{R}_a \implies $ $\mathbf{t}_i \sim \mathbf{t}_j$, so all the objects in $\mathcal{R}_a$ are connected in the undirected graph associated by $\sim$.
	\end{enumerate}
\end{proof}

\textbf{Sparsification of the directed graph:} Note that, in $G$, any directed edge from $\mathbf{t}_i$ to $\mathbf{t}_j$ such that $\mathbf{t}_i$ is at depth $k$ and $\mathbf{t}_j$ is at a depth greater than $k+1$ is not contained in the longest directed path to $\mathbf{t}_j$ in $G$.
Moreover, any modifications made to $G$ that do not impact the depths of its nodes have no influence on the calculated ranks according to Methodology~\ref{th:problem2}.
Bearing this in mind, we construct a graph $H$ from $G$ by removing all edges $(\mathbf{t}_i, \mathbf{t}_j)$ in $G$ where $d(\mathbf{t}_j) - d(\mathbf{t}_i) > 1$. The construction of $H$ ensures that the nodes at a specific depth are exclusively connected by directed edges only to nodes at the subsequent depth. 

\begin{figure}[h!]
	\centering
	\begin{subfigure}{0.48\textwidth}
		\centering
		\includegraphics[width=\textwidth]{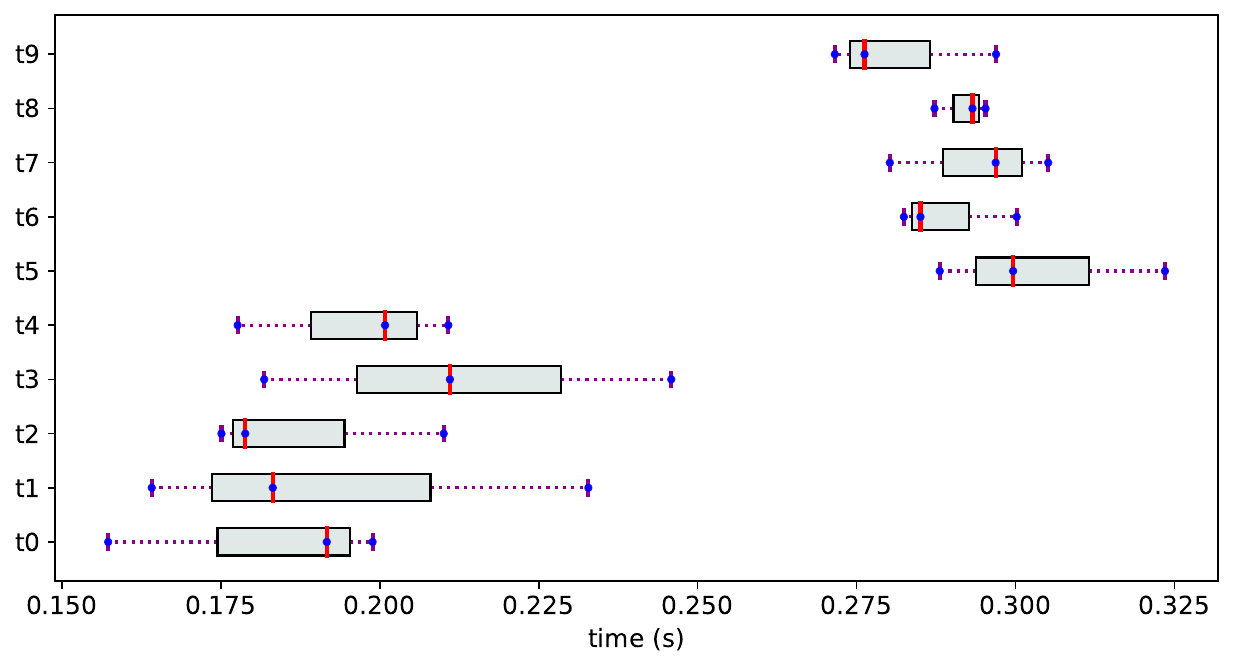}
		\caption{Sets of measurements $\mathcal{M}_4$. }
		\label{fig3:hasse-eg-2}
	\end{subfigure}
	\hfill
	\begin{subfigure}{0.48\textwidth}
		\centering
		\includegraphics[width=0.7\textwidth]{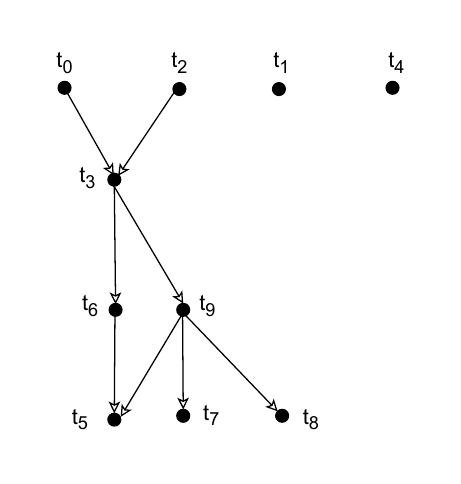}
		\caption{The graph $H$ associated with $\mathcal{M}_4$ and $<_{eg}$.\\$\mathcal{R}_0 = \{\mathbf{t}_0, \mathbf{t}_2, \mathbf{t}_1, \mathbf{t}_4\}$, $\mathcal{R}_1 = \{\mathbf{t}_3\}$, $\mathcal{R}_2 = \{\mathbf{t}_6, \mathbf{t}_9\}$, $\mathcal{R}_3 = \{\mathbf{t}_5, \mathbf{t}_7, \mathbf{t}_8\}$}
		\label{fig3:hasse-eg-2-dfg}
	\end{subfigure}
	\caption{An example to illustrate sparsification of $G$.  }
	\label{fig3:hasse-eg2}
\end{figure}

For example, consider the 10 variants consisting of measurements sampled from the following distribution functions :
\begin{itemize}
	\setlength{\itemsep}{0pt} 
	\item $\mathbf{t}_0, \mathbf{t}_1, \mathbf{t}_2, \mathbf{t}_3, \mathbf{t}_4 $: Sampled from $\mathcal{N}(0.2, 0.02)$
	\item $\mathbf{t}_5, \mathbf{t}_6, \mathbf{t}_7, \mathbf{t}_8, \mathbf{t}_9 $: Sampled from $\mathcal{N}(0.3, 0.02)$
\end{itemize}
For each variant, we sample 3 measurement values (with initialization seed set to 159) and prepare the sets of measurements $\mathcal{M}_4$ (shown in Fig.~\ref{fig3:hasse-eg-2}). The partial rankings computed according to $<_{eg}$ are shown in Fig.~\ref{fig3:hasse-eg-2-dfg}. Note that there is no path from $\mathbf{t}_4$ to $\mathbf{t}_9$ even though $\mathbf{t}_4 <_{eg} \mathbf{t}_9$ because the difference in the depths of $\mathbf{t}_4$ and $\mathbf{t}_9$ is 2.
Existence of the edge from $\mathbf{t}_4$ to $\mathbf{t}_9$ would not have changed the length of the longest directed path to $\mathbf{t}_9$.

\subsection{Methodology 2: Reduction in the number of ranks}

We now explain how the partial ranking produced by Methodology~\ref{th:problem2} (sparsified) can be modified to determine an alternate partial ranking that might reduce the total number of ranks. 
Given the graph $H$ constructed from $\mathcal{M}$ and $<_{\mathbf{P}}$, the  ordered set partition $\mathcal{R}_0, \dots, \mathcal{R}_k, \dots, \mathcal{R}_{K-1}$ consisting of $K$ ranks (where $K$ is the length of the longest directed path in $H$) is computed using Eq.~\ref{eq:rank-hesse}. In order to determine if it is possible to produce an alternate partial ranking with smaller $K$, we apply the following methodology:

\bigskip
\begin{mytheo}
	\label{th:problem3}
	Given the graph $H$ and the corresponding ordered set partition $\mathcal{R}_0,\dots, \mathcal{R}_{K-1}$, perform the following steps:
	
	\begin{enumerate}
		\item For all the nodes $\mathbf{t}_i$ in $H$, compute the number of incoming  and  outgoing edges.
		
		\item $\forall \mathcal{R}_k$, form a list\footnote{In order to be able to index the elements in the set $\mathcal{R}_k$, we introduce the list $\mathbf{R}_k$. The element at position $i$ (zero-based indexing) is denoted by $\mathbf{R}_k[i]$.} $\mathbf{R}_k$ by arranging the objects  in $\mathcal{R}_k$ according to decreasing number of outgoing edges in $H$. If there are objects with the same number of outgoing edges, arrange the objects with ties according to increasing number of incoming edges in $H$.
		
		\item Concatenate the lists into
		$\mathbf{T} = \mathbf{R}_0 \oplus \mathbf{R}_1 \oplus \ldots \oplus \mathbf{R}_{K-1}$. Note that the objects in $\mathbf{T}$ are arranged left to right from highest to lowest ranks.
		
		\item  Each $\mathbf{T}[i]$ is assigned  a rank $R[i]$  as follows: 
		\begin{enumerate}
			\item \textbf{set} $R[0] = 0$ (i.e., $\mathbf{T}[0]$ is assigned the best rank).
			\item \textbf{for} \textit{i} = $1,\dots N$:
			\begin{enumerate}
				\item \textbf{if} $\mathbf{T}[i-1]  <_{\mathbf{P}} \mathbf{T}[i]$ \textbf{then} \textbf{set} $R[i] = R[i-1] +1$.
				\item \textbf{else} \textbf{set} $R[i] = R[i-1]$.
			\end{enumerate}
		\end{enumerate}
	\end{enumerate}
\end{mytheo}
\bigskip
\noindent According to Step~4b of Methodology~\ref{th:problem3}, adjacent objects in $\mathbf{T}$ that are incomparable to one another are assigned the same rank. Let $\hat{K}$ be the number of unique values in $R$.
The ordered set partition of $\mathcal{M}$ consists of sets within  which all the objects are assigned the  same rank; i.e.,
\begin{equation}
	\label{eq:rank-sort}
	\mathcal{\hat{R}}_k = \{ \mathbf{T}[i] \in \mathbf{T}\ | \  R[i] = k \} \qquad \forall k \in \{0, \dots, \hat{K}-1\}
\end{equation}
and $\mathcal{\hat{R}}_k$ consists of objects that receive the rank $k$. The new number of ranks $\hat{K}$ is smaller than or equal to $K$.

\begin{table*}[h!]
	\centering
		\centering
		\caption{Partial ranking of $\mathcal{M}_4$ with $<_{eg}$ according to Methodology~\ref{th:problem3}.}
		\label{tab3:min-ranks-eg1}
		\begin{adjustbox}{width=.9\linewidth,center}
			\begin{tabular}{@{}ll cc cc cc cc cc cc cc cc cc cc@{}}
				\toprule
				$\mathbf{T}$&& $\mathbf{T}$[0] &&$\mathbf{T}$[1] && $\mathbf{T}$[2] && $\mathbf{T}$[3]&&$\mathbf{T}[4]$&&$\mathbf{T}[5]$&&$\mathbf{T}[6]$&&$\mathbf{T}[7]$&&$\mathbf{T}[8]$&&$\mathbf{T}[9]$&\\
				\cmidrule{3-22}
				&& $\mathbf{t}_0$ &$\sim$&  $\mathbf{t}_2$&$\sim$&  $\mathbf{t}_1$ &$\sim$ &  $\mathbf{t}_4$&$\sim$&$\mathbf{t}_3$&$<_{eg}$&$\mathbf{t}_9$&$\sim$&$\mathbf{t}_6$&$\sim$&$\mathbf{t}_7$&$\sim$&$\mathbf{t}_8$&$\sim$&$\mathbf{t}_5$& \\
				\midrule
				$R$ &&$R$[0] &&$R$[1] && $R$[2] && $R$[3]&&$R$[4]&&$R$[5]&&$R$[6]&&$R$[7]&&$R$[8]&&$R$[9]&\\
				\cmidrule{3-22}
				&& 0 &&  0 &&  0 &&   0&&0&&1&&1&&1&&1&&1& \\
				\midrule
				$\mathcal{\hat{R}}_k$&&\multicolumn{20}{c}{$\mathcal{\hat{R}}_0 = \{ \mathbf{t}_0, \mathbf{t}_1, \mathbf{t}_2, \mathbf{t}_3, \mathbf{t}_4\}$, $\mathcal{\hat{R}}_1 = \{\mathbf{t}_5, \mathbf{t}_6, \mathbf{t}_7, \mathbf{t}_8, \mathbf{t}_9\}$}\\
				\bottomrule
			\end{tabular}
		\end{adjustbox}

\end{table*}

\textbf{Illustrative Example}:  Consider the graph $H$ constructed from the sets of measurements $\mathcal{M}_4$ ranked according to $<_{eg}$  as shown in Fig.~\ref{fig3:hasse-eg2}. The ordered set partition $\mathcal{R}_0, \mathcal{R}_1, \mathcal{R}_2, \mathcal{R}_3$ is shown in Fig.~\ref{fig3:hasse-eg-2-dfg}. The arrangements according to Step~2 of Methodology~\ref{th:problem3} are:
\begin{itemize}
	\setlength{\itemsep}{0pt} 
	\item $\mathbf{R}_0 = [\mathbf{t}_0, \mathbf{t}_2, \mathbf{t}_1, \mathbf{t}_4]$
	\item $\mathbf{R}_1 = [\mathbf{t}_3]$
	\item $\mathbf{R}_2 = [\mathbf{t}_9, \mathbf{t}_6]$
	\item $\mathbf{R_3 = [\mathbf{t}_7, \mathbf{t}_8, \mathbf{t}_5]}$
\end{itemize}
The concatenated list according to Step~3 is:
\begin{equation}
	\label{eq3:inp-arrange}
	\mathbf{T} = [\mathbf{t}_0, \mathbf{t}_2, \mathbf{t}_1, \mathbf{t}_4, \mathbf{t}_3, \mathbf{t}_9, \mathbf{t}_6, \mathbf{t}_7, \mathbf{t}_8, \mathbf{t}_5]
\end{equation}
The ranking computed according to Methodology~\ref{th:problem3} with the list $\mathbf{T}$ in Eq.~\ref{eq3:inp-arrange} is shown in Table~\ref{tab3:min-ranks-eg1}, and consists of only two ranks.
\begin{figure}[h!]
		\centering
		\includegraphics[width=0.5\textwidth]{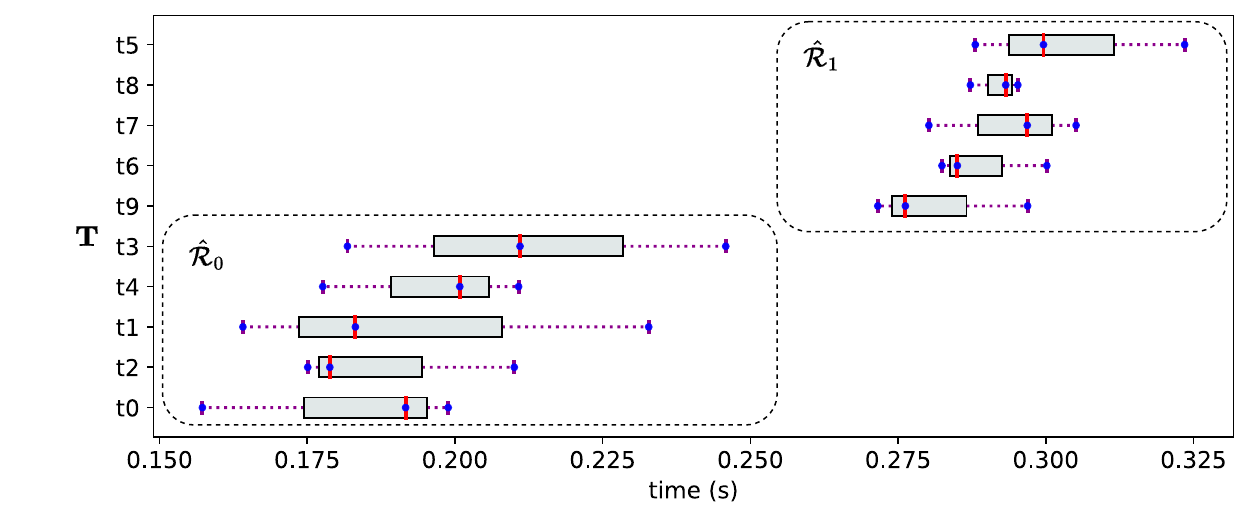}
		\caption{The annotation of the partial ranking on $\mathcal{M}_4$  with $<_{eg}$ according to Methodology~\ref{th:problem3}. }
		\label{fig3:hasse-eg2-r}
\end{figure}

In the step 2 of Methodology~\ref{th:problem3},  if $\exists \mathbf{t}_i \in \mathcal{R}_k$ and $\exists \mathbf{t}_j \in \mathcal{R}_{k+1}$ such that $\mathbf{t}_i \sim \mathbf{t}_j$, then $\mathbf{t}_i$ is pushed towards the right in the list $\mathbf{R}_k$, and $\mathbf{t}_j$ is pushed towards the left in the list $\mathbf{R}_{k+1}$, and then, the ranks $\forall \mathbf{t}_i \in \mathcal{R}_k$ and $\forall \mathbf{t}_j \in \mathcal{R}_{k+1}$ could be merged according to Step~4. For illustration, 
let us denote the last element in a list $\mathbf{R}_k$ as $\mathbf{R}_k[-1]$. In our example, as $\mathbf{R}_0[-1] = \mathbf{t}_4 \sim \mathbf{t}_3 = \mathbf{R}_1[0] $,  the concatenation of $\mathbf{R}_0$ and $\mathbf{R}_1$ merges the ranks of the variants in $\mathcal{R}_0$ and $\mathcal{R}_1$. Similarly, as $\mathbf{R}_2[-1] = \mathbf{t}_6 \sim \mathbf{t}_7 = \mathbf{R}_3[0]$, the ranks of the variants in $\mathcal{R}_2$ and $\mathcal{R}_3$ are also merged. For the sake of clarity, the sets of measurements $\mathcal{M}_4$ are shown again in Fig.~\ref{fig3:hasse-eg2-r}, but now with the y-axis re-arranged from bottom to top based on the position of the variants in the list $\mathbf{T}$, starting from $\mathbf{T}[0]$, and the ranking is annotated. 
\bigskip
\begin{mylem}
	\label{th:lemma1}
	Rankings produced by Methodology~\ref{th:problem3} are partial rankings (Def.~\ref{th:ranking}).
\end{mylem}
\bigskip
\begin{proof}
	We need to prove that the ranking follows the three properties in Definition~\ref{th:ranking}.
	\begin{enumerate}
		\item Proof for Property 1:
		\begin{itemize}
			\item $\forall \mathbf{T}[i] \in \mathcal{\hat{R}}_a$ and  $\forall \mathbf{T}[j] \in \mathcal{\hat{R}}_b$, $a<b \implies R[i] < R[j]$ (Eq.~\ref{eq:rank-sort}).
			\item $R[i] < R[j] \implies i<j$  (Meth.~\ref{th:problem3}: Step 4b)
			\item $i<j \implies \mathbf{T}[i] <_{\mathbf{P}} \mathbf{T}[j]$ or $\mathbf{T}[i] \sim \mathbf{T}[j]$ (Meth.~\ref{th:problem3}: Step 3; the objects in $\mathbf{T}$ are arranged from left to right from highest to lowest ranks computed according to Methodology~\ref{th:problem2}).
		\end{itemize}
		Hence, $\forall \mathbf{T}[i] \in \mathcal{\hat{R}}_a$ and  $\forall \mathbf{T}[j] \in \mathcal{\hat{R}}_b$, $a<b \implies\mathbf{T}[i] <_{\mathbf{P}} \mathbf{T}[j]$ or $\mathbf{T}[i] \sim \mathbf{T}[j]$.
		
		\item Proof for Property 2: According to Meth.~\ref{th:problem3}: Step 4b(i), a new rank is created only when there exists $i$ such that $\mathbf{T}[i-1]  <_{\mathbf{P}} \mathbf{T}[i]$. Hence, for every pair of ranks $\mathcal{\hat{R}}_a, \mathcal{\hat{R}}_b$, it is possible to find a pair $\mathbf{t}_i \in \mathcal{\hat{R}}_a$ and $\mathbf{t}_j \in \mathcal{\hat{R}}_b$ such that $\mathbf{t}_i <_{\mathbf{P}} \mathbf{t}_j$.   
		
		
		\item Proof for Property 3: $|\mathcal{\hat{R}}_a| > 1 \implies $   $\exists k,l$ such that $\mathbf{T}[k] \sim \mathbf{T}[k+1] \sim \dots \sim  \mathbf{T}[l]$, and all the objects from positions $k$ to $l$ in $\mathbf{T}$ are  the only objects in $\mathcal{\hat{R}}_a$ (Meth.~\ref{th:problem3}: Step 4). 
		Hence,  there exists an arrangement of $\mathcal{\hat{R}}_a$ where the adjacent objects are pair-wise incomparable. This means that the objects in $\mathcal{\hat{R}}_a$ are connected in the undirected graph associated by $\sim$. 
	\end{enumerate}
\end{proof}

The number of ranks computed by Methodology~\ref{th:problem3} is either smaller than or equal to the number of ranks computed according to Methodology~\ref{th:problem2}, but not necessarily the least possible number of ranks.


\begin{figure}[h!]
	\centering
	\begin{subfigure}{0.5\textwidth}
		\centering
		\includegraphics[width=\textwidth]{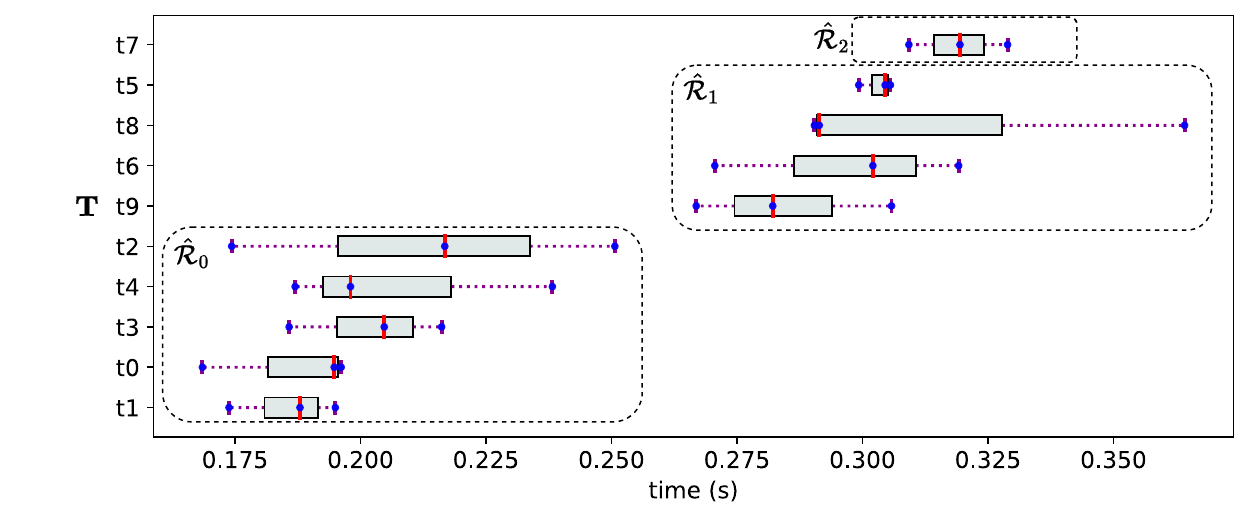}
		\caption{Methodology~\ref{th:problem3} on $\mathcal{M}_5$. }
		\label{fig3:minrank-eg4-r}
	\end{subfigure}
	\hfill
	\begin{subfigure}{0.5\textwidth}
		\centering
		\includegraphics[width=\textwidth]{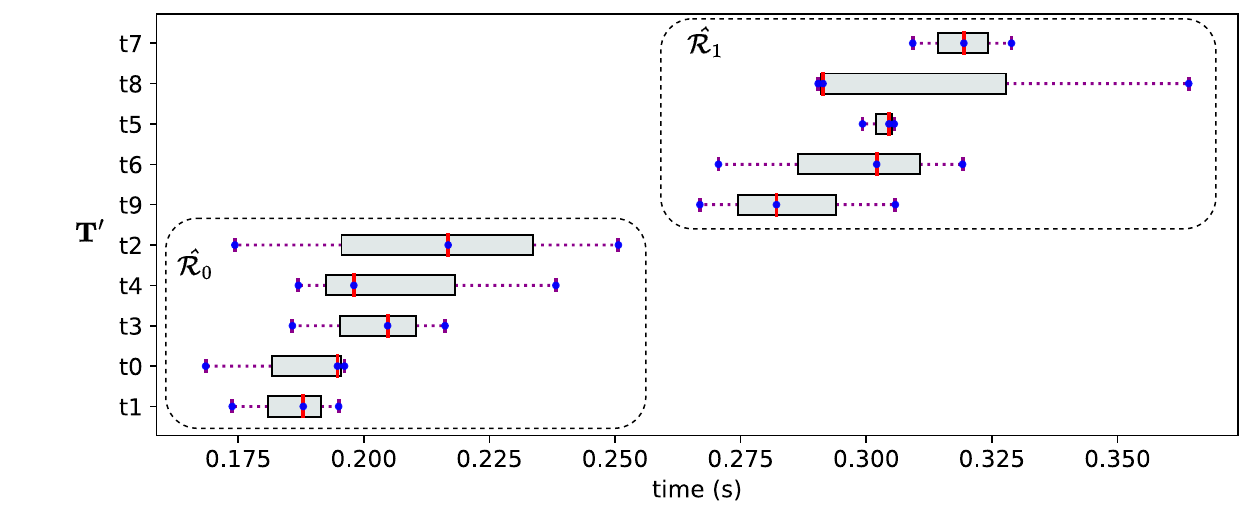}
		\caption{Methodology~\ref{th:problem3} on $\mathcal{M}_5$ with arrangement $\mathbf{T'}$.}
		\label{fig3:minrank-eg4-c}
	\end{subfigure}
	\hfill
	\begin{subfigure}{0.45\textwidth}
		\centering
		\includegraphics[width=0.95\textwidth]{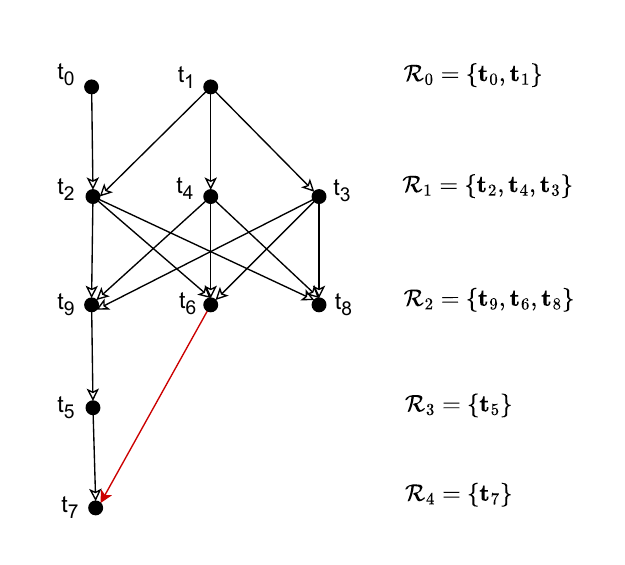}
		\caption{Methodology~\ref{th:problem2} on $\mathcal{M}_5$.  The transitivity reduction of graph $G$ for $\mathcal{M}_5$ according to $<_{eg}$. By removing the edge indicated in red, the graph $H$ is obtained.}
		\label{fig3:minrank-eg4-dfg}
	\end{subfigure}
	\caption{Partial ranking of $\mathcal{M}_5$ with $<_{eg}$. An illustrative example to demonstrate that Methodology~\ref{th:problem3} does not always find the partial ranking with the least number of ranks.  }
	\label{fig3:minrank-eg4}
\end{figure}

\textbf{Instances where the least number of ranks is not computed}:
In certain cases, Methodology~\ref{th:problem3} does not yield a partial ranking with the minimum possible number of ranks. For example, consider the sets of measurements $\mathcal{M}_5$ shown in Fig.~\ref{fig3:minrank-eg4}. Given $\mathcal{M}_5$ and $<_{eg}$, the arrangement $\mathbf{T}$ and the annotations of the partial ranking calculated according to Methodology~\ref{th:problem3} are shown in Fig.~\ref{fig3:minrank-eg4-r}. While this partial ranking consists of three ranks, it is possible to form an alternate arrangement $\mathbf{T}'$ (shown in Fig.~\ref{fig3:minrank-eg4-c}), that produces a partial ranking with just two ranks according to Methodology~\ref{th:problem3} (Step~4).

A sufficient condition for which Methodology~\ref{th:problem3} will not produce a partial ranking with the least number of ranks is when $\exists \mathbf{t}_i \in \mathcal{R}_k$ and $\exists \mathbf{t}_j \in \mathcal{R}_{k+2}$ such that $\mathbf{t}_i \sim \mathbf{t}_j$, but $\nexists \mathbf{t}_m \in \mathcal{R}_{k+1}$ such that $\mathbf{t}_m \sim \mathbf{t}_j$.
For example, in $\mathcal{M}_5$, the directed graph $G$ (with transitivity reduction) is shown in Fig.~\ref{fig3:minrank-eg4-dfg} (by removing the edge highlighted in red, the graph $H$ is obtained). According to Methodology~\ref{th:problem2}, the number of ranks is 5, and we just saw that it possible to have a partial ranking consisting of 2 ranks. Methodology~\ref{th:problem3} cannot reduce the number of ranks to smaller than 3 because $\exists \mathbf{t}_8 \in \mathcal{R}_2$ and $\exists \mathbf{t}_7 \in \mathcal{R}_4$ such that $\mathbf{t}_8 \sim \mathbf{t}_7$, but  $\nexists \mathbf{t}_m \in \mathcal{R}_3$ such that $\mathbf{t}_m \sim \mathbf{t}_7$. As a consequence, according to Methodology~\ref{th:problem3} (Step~3), $\mathbf{t}_7 \in \mathbf{R}_4$ cannot be placed next to $\mathbf{t}_8 \in \mathbf{R}_2$ because $\mathbf{t}_5 \in \mathbf{R}_3$ has to be placed between $\mathbf{t}_8$ and $\mathbf{t}_7$, and hence the ranks of $\mathbf{t}_8$ and $\mathbf{t}_7$ cannot be merged.


\subsection{Methodology 3: For minimum number of ranks}


We now explain a methodology for computing the partial ranking with the minimum number of ranks. For a given $\mathcal{M}$ and $<_{\mathbf{P}}$, let $U$ be an undirected graph such that $\mathbf{t}_i \in \mathcal{M}$ are the nodes and an edge between $\mathbf{t}_i$ and $\mathbf{t}_j$ exists if and only if $\mathbf{t}_i \sim \mathbf{t}_j$ according to $<_{\mathbf{P}}$.
\bigskip
\begin{mytheo}
	\label{th:problem4}
	Given the graph $U$ constructed from the partial order $(\mathcal{M}, <_{\mathbf{P}})$,
	\begin{enumerate}
		\item Partition $U$ to create $\mathcal{U} = \{\mathcal{V}_0, \dots, \mathcal{V}_{K-1}\}$ such that each partition $\mathcal{V}_k \in \mathcal{U}$ corresponds to the connected components of $U$. 
		\item Construct a directed graph $G'$ such that $\mathcal{V}_i \in \mathcal{U}$ are the nodes and an edge from $\mathcal{V}_i$ to $\mathcal{V}_j$ exists if and only if for any $\mathbf{t}_k \in \mathcal{V}_i$ and  $\mathbf{t}_l \in \mathcal{V}_j$, $\mathbf{t}_k <_{\mathbf{P}} \mathbf{t}_l$. 
		\item $\forall \mathcal{V}_i \in \mathcal{U}$, set the rank of all $\mathbf{t}_k \in \mathcal{V}_i$ equal to the depth of $\mathcal{V}_i$ in $G'$.
	\end{enumerate}
\end{mytheo}
\bigskip

\begin{figure}[h!]
	\centering
	\begin{subfigure}{0.45\textwidth}
		\centering
		\includegraphics[width=\textwidth]{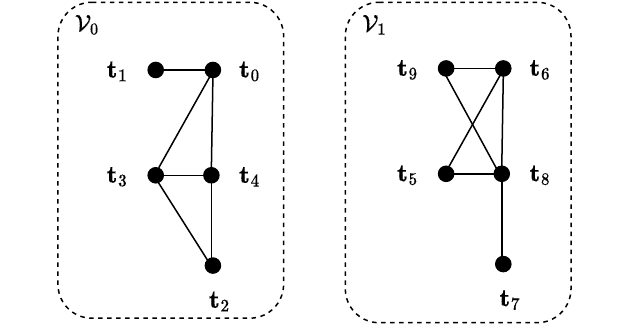}
		\caption{The undirected graph $U$ and the connected components $\mathcal{V}_0$ and $\mathcal{V}_1$. }
		\label{fig3:met4-eg-udg}
	\end{subfigure}
	\hfill
	\begin{subfigure}{0.35\textwidth}
		\centering
		\includegraphics[width=\textwidth]{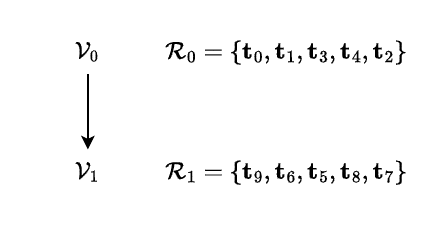}
		\caption{The graph $G'$ and the partial ranks.}
		\label{fig3:met4-eg-dfg}
	\end{subfigure}
	\caption{Methodology~\ref{th:problem4} on $\mathcal{M}_5$ and $<_{eg}$.  }
	\label{fig3:met4-eg}
\end{figure}

\textbf{Illustrative example:} Consider again the sets of measurements $\mathcal{M}_5$ and the relation $<_{eg}$. According to Step~1, the undirected graph $U$ consists of two partitions $\mathcal{V}_0$ and $\mathcal{V}_1$ (shown in Fig.~\ref{fig3:met4-eg-udg}). The directed graph $G'$ and the resulting ordered set partitioning are shown in Fig.~\ref{fig3:met4-eg-dfg}.   

\begin{mylem}
	\label{th:lemma4}
	Rankings produced by Methodology~\ref{th:problem4} are partial rankings (Def.~\ref{th:ranking}) and consists of minimum possible number of ranks.
\end{mylem}
\begin{proof}
	Notice that $\forall \mathcal{R}_a, \mathcal{R}_b$, $a<b \implies \forall \mathbf{t}_k \in \mathcal{R}_a$ and $ \forall \mathbf{t}_l \in \mathcal{R}_b$, $\mathbf{t}_k <_{\mathbf{P}} \mathbf{t}_l$. Thus, the requirements according to Property~1 and Property~2 are satisfied. The condition according to Property~3 is naturally satisfied as all the variants in a particular partition $\mathcal{V}_i$ are connected by the incomparability relation $\sim$. Hence, the ranking computed according to Methodology~\ref{th:problem4} is a partial ranking.
	
	We prove by contradiction that the partial ranking according to Methodology~\ref{th:problem4} consists of the minimum number of ranks. Suppose that there exists some partial ranking with strictly fewer than $K$ ranks. Then, by the pigeonhole principle, at least one rank must contain a pair of variants that are disconnected in the graph associated with the incomparability relation. This violates Property 3 and hence there cannot exist any partial ranking with fewer than $K$ ranks.
\end{proof}

\begin{figure}[b!]
	\centering
	\begin{subfigure}{0.35\textwidth}
		\centering
		\includegraphics[width=\textwidth]{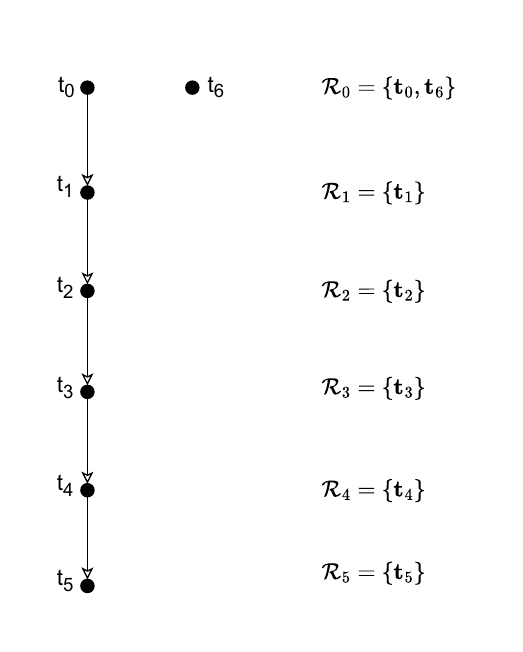}
		\caption{Methodology~\ref{th:problem2} on $\mathcal{M}_6$.}
		\label{fig3:minrank-eg3-dfg}
	\end{subfigure}
	
	\hfill
	
	\begin{subfigure}{0.5\textwidth}
		\centering
		\includegraphics[width=\textwidth]{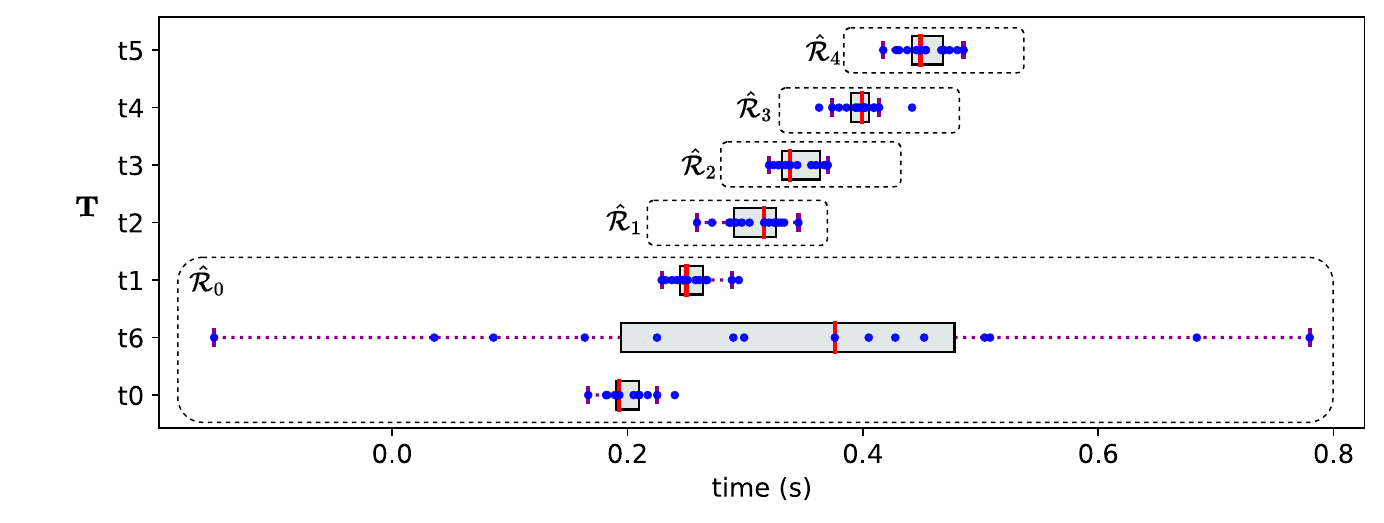}
		\caption{Methodology~\ref{th:problem3} on $\mathcal{M}_6$.}
		\label{fig3:minrank-eg3-r}
	\end{subfigure}
	
	\caption{Partial ranking of $\mathcal{M}_6$ with $<_{eg}$.  }
	\label{fig3:minrank-eg3}
\end{figure}

\subsection{Implications on some special cases:} 
When there are numerous variants with their measurement intervals distinctly separated from each other, except for a single variant whose interval overlaps with all the others, Methodology~\ref{th:problem4} merges all the variants into one rank. Such a ranking is sometimes undesirable because it is not possible to discriminate between any of the distinctly separated variants.
For example, consider the sets of measurements $\mathcal{M}_6$ shown in Fig.~\ref{fig3:minrank-eg3-r}. Each variant consists of 15 measurement values and the IQI of all the variants except for $\mathbf{t}_6$ are distinctly separated from one another. 
For $\mathcal{M}_6$, Methodology~\ref{th:problem2} produces a partial ranking with six ranks as shown in Fig.~\ref{fig3:minrank-eg3-dfg}, and this is the maximum number of ranks that could be produced for $\mathcal{M}_6$ with $<_{eg}$; $\mathbf{t}_0, \mathbf{t}_6 \in \mathcal{R}_0$ and all the other variants in distinct ranks. Methodology~\ref{th:problem3} reduces the number of ranks by 1, and the resulting partial rankings are annotated in Fig.~\ref{fig3:minrank-eg3-r}.
Methodology~\ref{th:problem4} however, produces a partial ranking where all the variants are placed in a single rank. 

\begin{figure}[h!]
	\centering
	\begin{subfigure}{0.45\textwidth}
		\centering
		\includegraphics[width=\textwidth]{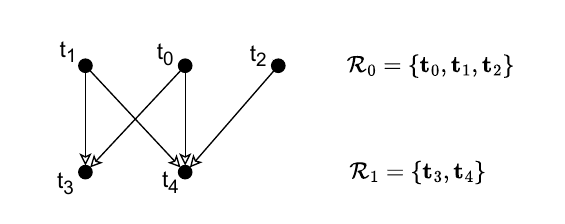}
		\caption{Methodology~\ref{th:problem2} on $\mathcal{M}_7$.}
		\label{fig3:minrank-eg5-dfg}
	\end{subfigure}
	\hfill
	\begin{subfigure}{0.5\textwidth}
		\centering
		\includegraphics[width=\textwidth]{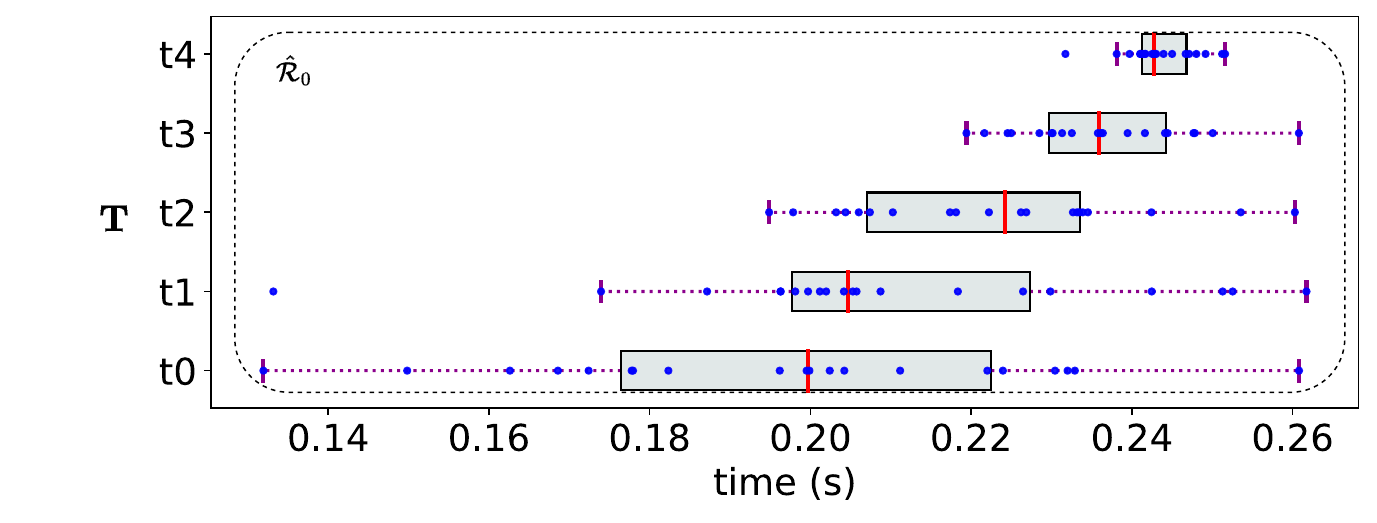}
		\caption{Methodology~\ref{th:problem3} on $\mathcal{M}_7$.}
		\label{fig3:minrank-eg5-r}
	\end{subfigure}
	\caption{Partial ranking of $\mathcal{M}_7$ with $<_{eg}$.  }
	\label{fig3:minrank-eg5}
\end{figure}

Let us examine one more example that showcases variants with overlapping yet monotonically increasing differences, such as the sets of measurements $\mathcal{M}_7$ depicted in Fig.~\ref{fig3:minrank-eg5}. In this case, although assigning a distinct rank to each variant seems more intuitive\footnote{In this context, the way in which a ranking intuitively reflects the information present in the data.}, both Methodology~\ref{th:problem3} and Methodology~\ref{th:problem4} produces a partial ranking where all the variants are placed in the same rank (as shown in Fig.~\ref{fig3:minrank-eg5-r}). Methodology~\ref{th:problem2} computes a partial ranking with 2 ranks (as shown in Fig.~\ref{fig3:minrank-eg5-dfg}). Notice that Methodology~\ref{th:problem2} does not compute the maximum possible number of ranks, because for $\mathcal{M}_7$, the following partial ranking with 3 ranks exists: $\mathcal{R}_0 = \{\mathbf{t}_0, \mathbf{t}_1\}$,  $\mathcal{R}_1 = \{\mathbf{t}_2, \mathbf{t}_3\}$ and  $\mathcal{R}_2 = \{\mathbf{t}_4\}$.

\section{Handling the effects of the Quantile Intervals }
\label{sec3:handlingq}
\begin{figure}[b!]
	\centering
	\includegraphics[width=\linewidth]{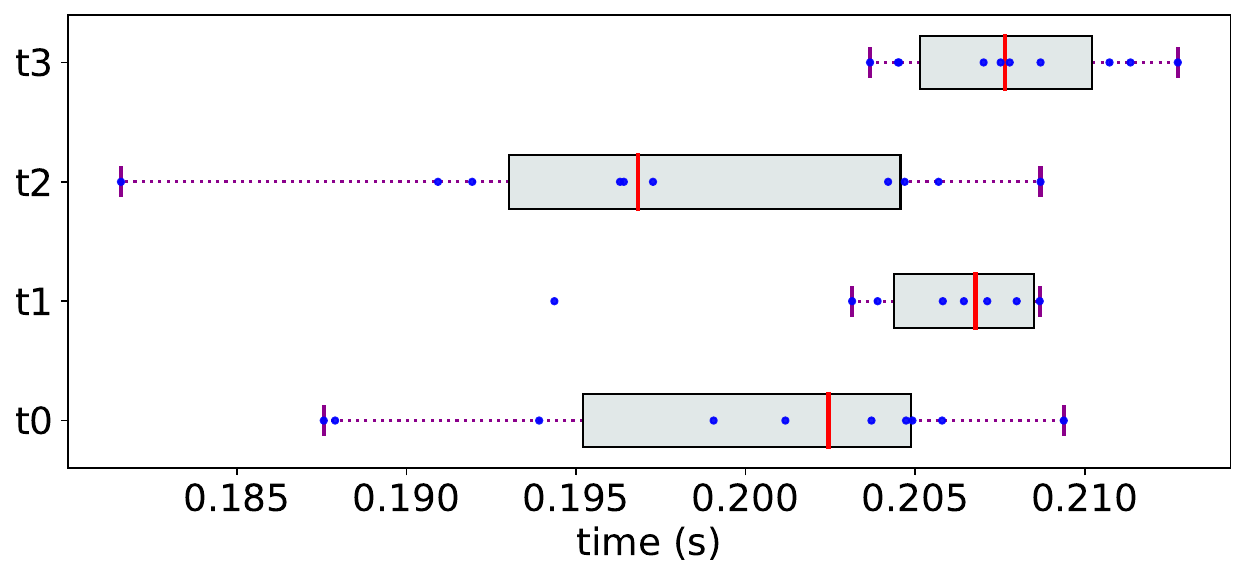}
	\captionof{figure}{Sets of measurements $\mathcal{M}_8$.}
	\label{fig:rel-eg}
\end{figure}

Thus far, we have utilized the relation $<_{eg}$ to compare two sets of measurements, which makes comparisons based on overlap in the Inter Quartile Intervals (IQIs), or in other words,  based on overlap in the Inter \textit{Quantile} Interval (IQnI) between the 25th and 75th quantiles among the variants. Sometimes, altering the quantiles can lead to a different ranking. 
For instance, consider the sets of measurements $\mathcal{M}_8$ shown in Fig.~\ref{fig:rel-eg} where it could be intuitive for one to expect the following ranking:
\begin{equation}
	\label{eq:rel-rank}
	\mathcal{R}_0 =\{ \mathbf{t}_0, \mathbf{t}_2 \}, \quad \mathcal{R}_0 =\{ \mathbf{t}_1, \mathbf{t}_3 \}.
\end{equation}
However, the IQnI between the 25th and the 75th quantile  of $\mathbf{t}_1$ \textit{slightly} overlaps with that of $\mathbf{t}_0$ and $\mathbf{t}_2$. As a consequence, when using the relation $<_{eg}$ to compare the variants,  $\mathbf{t}_0$, $\mathbf{t}_1$ and $\mathbf{t}_2$ are all pairwise incomparable and obtain the same rank.
In order to alleviate this problem, for a given ranking methodology, we compute multiple rankings corresponding to different quantiles, and average them. In what follows, we propose a systematic way of choosing a ranking corresponding to one among several quantile intervals based on the averaged or the mean ranks. 

\begin{table*}[h!]
	\centering
	\caption{Reliability Scores of the ranks computed by Methodology~\ref{th:problem2} on $\mathcal{M}_8$ with $<_{(l,u)}$ at different quantile limits.}
	\label{tab:rel}
	\begin{adjustbox}{width=0.6\textwidth,center}
		\begin{tabular}{@{}ll ccccccccc@{}}
			\toprule
			&&$\mathbf{t}_0$ &&$\mathbf{t}_1$ &&$\mathbf{t}_2$ &&$\mathbf{t}_3$ &&avg\_rel$(l,u)$ \\
			\midrule
			$r(\mathbf{t}_i,25,75)$&& 0&&0&&0&&1&&-\\
			$r(\mathbf{t}_i,30,70)$&& 0&&1&&0&&1&&-\\
			$r(\mathbf{t}_i,35,65)$&& 0&&1&&0&&1&&-\\
			$r(\mathbf{t}_i,40,60)$&& 1&&2&&0&&2&&-\\
			\midrule
			$mr(\mathbf{t}_i)$ &&0.25&&1.00&&0.00&&1.25&&-\\
			\midrule
			$rel(\mathbf{t}_i,25,75)$&& -0.25&&-1.0&&0.0&&-0.25&&\textbf{-0.38}\\
			$rel(\mathbf{t}_i,30,70)$&& -0.25&&-0.00&&0.0&&-0.25&&\textbf{-0.13}\\
			$rel(\mathbf{t}_i,35,65)$&& -0.25&&-0.00&&0.0&&-0.25&&\textbf{-0.13}\\
			$rel(\mathbf{t}_i,40,60)$&& -0.75&&-1.00&&0.0&&-0.75&&\textbf{-0.63}\\
			\bottomrule
		\end{tabular}
	\end{adjustbox}

\end{table*}

To this end, let us first generalize $<_{eg}$ to accommodate arbitrary quantile limits. Let $l$ and $u$ denote  the lower and the upper quantile limit respectively. For each set of measurements values $\mathbf{t}_i \in \mathbb{R}^{M}$,  sort $\mathbf{t}_i$ in the ascending order and let $t_i^{l}, t_i^{u} \in \mathbb{R}$ be the linearly interpolated measurement values at the $l$ and $u$ quantiles respectively. Then, for each $\mathbf{t}_i$, $(t_i^{l}, t_i^{u})$ is the IQnI between the quantile limit $(l,u)$. We define the relation  $<_{(l,u)}$ as:

\begin{definition}
	\label{th:compare}
	For a given $(l, u)$, $\mathbf{t}_i <_{(l,u)} \mathbf{t}_j$ if and only if $t_i^{u} < t_j^{l}$.
\end{definition}
\bigskip

\noindent  Thus, inferring from Definition~\ref{th:compare}, two sets of measurements are considered incomparable if and only if their IQnIs overlap with one another. If $l = 25$  and $u = 75$, then IQnI becomes the IQI indicated in our box plots; consequently, $<_{eg}$ is same as $<_{(25,75)}$.

Let $\mathcal{Q}$ be a set of quantile limits. For a given ranking methodology, let $r(\mathbf{t}_i, l, u)$ be the rank calculated for the variant $\mathbf{t}_i \in \mathcal{M}$ at a particular $(l, u) \in \mathcal{Q}$. As the choice of ($l, u$) influences the partial rank calculation, instead of computing a partial ranking at just one $(l,u)$, we compute the partial ranks at several quantile limits, i.e., $\forall (l,u) \in \mathcal{Q}$, and calculate the mean rank of each variant. Let the mean rank of $\mathbf{t}_i$ be denoted by $mr(\mathbf{t}_i)$. Then, the reliability score $rel(\mathbf{t}_i, l,u)$ of the rank assigned to the variant $\mathbf{t}_i$ at a particular quantile limit $(l,u)$ is defined as:
\begin{equation}
	\label{eq:rel-eq}
	rel(\mathbf{t}_i, l, u) = -|r(\mathbf{t}_i,l, u) - mr(\mathbf{t}_i)|
\end{equation}
where a higher reliability score can be considered as a better ranking for the given measurement data, a better-than relation and $\mathcal{Q}$. That is, for the variant $\mathbf{t}_i$, the lower the difference of the rank $r(\mathbf{t}_i, l,u)$ calculated at a particular $(l,u)$ from the mean of the ranks $mr(\mathbf{t}_i)$  calculated at several quantile limits, the higher the reliability score  $rel(\mathbf{t}_i, l, u)$, indicating that the rank assignment $r(\mathbf{t}_i, l,u)$ is more reliable. The average reliability (avg\_rel) of a partial ranking at a quantile limit $(l,u)$ is the mean of the reliability scores at that $(l,u)$ from all the variants; that is,
\begin{equation}
	\text{avg\_rel}(l,u) = \frac{\sum\limits_{\mathbf{t}_i} rel(\mathbf{t}_i, l, u)}{|\mathcal{M}|} 
\end{equation}
where $|\mathcal{M}|$ is the total number of objects.

Thus, given $\mathcal{Q}$, we propose to compute partial rankings and average reliability scores $\forall (l,u) \in \mathcal{Q}$, and select a ranking with the highest average reliability score. 

\textbf{Illustrative example:} Let us compute partial rankings for $\mathcal{M}_8$ using Methodology~\ref{th:problem2} at all
\begin{equation}
	\label{eq:qlist}
	(l, u) \in  \{ (25,75), (30,70), (35,65), (40,60) \}.
\end{equation}
The mean ranks of the variants, the reliability scores of the rank assignments and the average reliability scores are shown in Table~\ref{tab:rel}.   The partial rankings at the limits $(30,70)$ and $(35, 65)$, which classifies the variants into ranks as expected according to Eq.~\ref{eq:rel-rank}, obtains a higher average reliability score than the partial rankings at other limits. 

\section{Experiments: Mining for the causes of performance differences from the measurement data}  
\label{sec3:exp}

\begin{table*}[h!]
	\begin{minipage}{1\textwidth}
		\centering
		\includegraphics[width=0.55\linewidth]{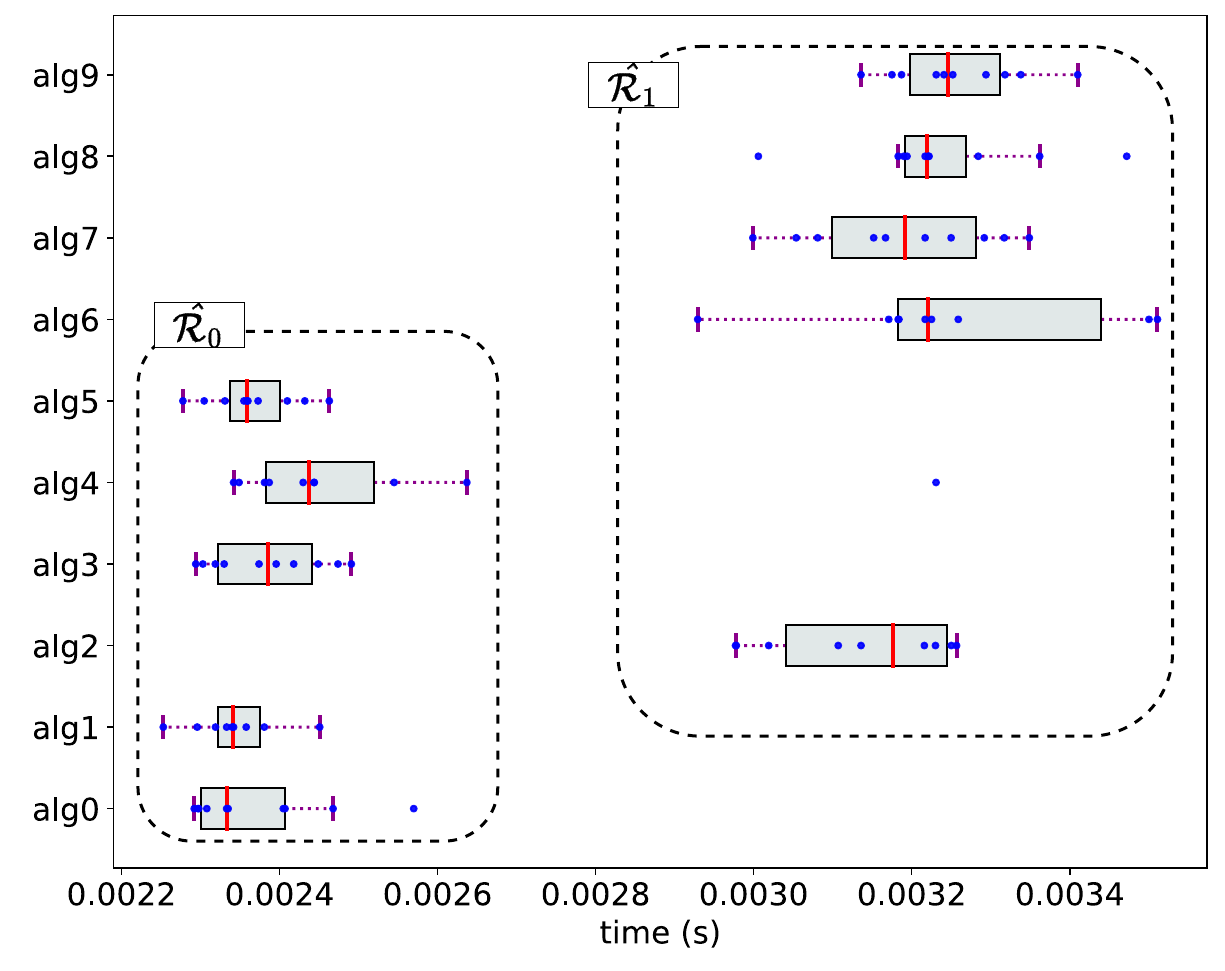}
		\captionof{figure}{The  time measurements of the algorithmic variants of the GLS problem:  $(X^{T}M^{-1}X)^{-1}X^{T}M^{-1}\mathbf{y}$ where  $X \in \mathbb{R}^{1000 \times 100}$, $M \in \mathbb{R}^{1000 \times 1000}$ and $\mathbf{y} \in \mathbb{R}^{1000}$ ($\mathcal{M}_{gls}$). The partial rankings using Methodology~\ref{th:problem3} according to $<_{eg}$ are annotated.}
		\label{fig3:gls-eg}
	\end{minipage}
	\hfill
	
	\begin{minipage}{1\textwidth}
		\centering
		\captionof{table}{Kernel sequences of the ten algorithmic variants considered in Fig.~\ref{fig3:gls-eg}.}
		\label{tab3:gls-seq}
		\begin{adjustbox}{width=0.65\columnwidth,center}
			\begin{tabular}{@{}l cc@{}}
				\toprule
				\textbf{Variant} &&\textbf{Kernel sequence}\\
				\midrule
				$\mathbf{alg}_0$&& potrf, trsm, trsv, syrk, gemv, potrf, trsv, trsv \\
				$\mathbf{alg}_1$&& potrf, trsv, trsm, syrk, potrf, gemv, trsv, trsv\\
				$\mathbf{alg}_{2}$&& potrf, trsm, trsv, syrk, gemv, qr, gemv, trsv\\
				$\mathbf{alg}_{3}$&& potrf, trsv, trsm, gemm, potrf, gemv, trsv, trsv \\
				$\mathbf{alg}_{4}$&& potrf, trsm, trsv, gemm, gemv, potrf, trsv, trsv \\
				$\mathbf{alg}_{5}$&&  potrf, trsm, trsv, gemm, potrf, gemv, trsv, trsv \\
				$\mathbf{alg}_{6}$&& transpose, potrf, trsm, trsv, syrk, potrf, trsv, trsm, gemv, trsv\\
				$\mathbf{alg}_{7}$&& transpose, potrf, trsm, syrk, potrf, trsv, trsv, trsm, gemv, trsv \\
				$\mathbf{alg}_{8}$&& transpose, potrf, trsm, syrk, potrf, trsm, trsm, trsv, trsv, gemv \\
				$\mathbf{alg}_{9}$&&  transpose, potrf, trsm, syrk, potrf, trsv, trsv, trsm, trsm, gemv \\
				\bottomrule
			\end{tabular}
		\end{adjustbox}

	\end{minipage}
	
\end{table*}

Partial ranking of a set of objects can be seen as clustering the objects into performance classes; i.e., a clustering in which there is a notion of one cluster being better than another.
In this section, we demonstrate that identifying performance classes using our partial ranking methodologies automate the discovery of the causes of performance differences between the variants. For our experiments, we revisit the execution time measurements of the algorithmic variants of the Generalized Least Square (GLS) problem introduced in Sec.~\ref{sec3:int}, and identify the library calls that impact performance the most (Sec.~\ref{sec3:gls}). Then,  we apply our partial ranking methodology on a real-life dataset from a Business Process application to identify the underlying causes of inefficiencies (Sec.~\ref{sec3:bpi}).

\subsection{Algorithmic Variants of GLS}
\label{sec3:gls}

The execution time measurements of ten algorithmic variants of GLS for given matrix dimensions are shown again in Fig.~\ref{fig3:gls-eg}. Each algorithmic variant can be identified as a sequence of library or kernel calls provided by optimized libraries such as BLAS and LAPACK. The sequences of kernel calls for the ten algorithmic variants as generated by the Linnea compiler~\cite{barthels2021linnea} are shown in Table~\ref{tab3:gls-seq}.
Let us first attempt to elucidate the causes of performance differences among the variants in terms of the kernel calls, solely by looking at the data $\mathcal{M}_{gls}$, without relying on any specific ranking methodology. At first glance, one could intuitively categorize the variants into two distinct groups: Fast variants  and Slow variants, as represented  below:
\begin{equation}
	\label{eq3:gls-gt}
	\begin{aligned}
		\text{Fast} &= \{\mathbf{alg}_0, \mathbf{alg}_1, \mathbf{alg}_3, \mathbf{alg}_4, \mathbf{alg}_5\} \\
		\text{Slow} &= \{\mathbf{alg}_2, \mathbf{alg}_6, \mathbf{alg}_7, \mathbf{alg}_8, \mathbf{alg}_9\}
	\end{aligned}
\end{equation}

By carefully observing the similarities and differences in the kernel sequences among the variants in the Fast and Slow groups based on the data $\mathcal{M}_{gls}$, one could discern the following associations:
\begin{enumerate}
	\itemsep0em 
	\item[\textbf{r1)}] Only the fast variants make use of the kernel \textit{gemm}.
	\item[\textbf{r2)}] Only the slow variants make use of the kernel \textit{transpose}.
	\item[\textbf{r3)}] Only the slow variants make use of the kernel \textit{qr}.
\end{enumerate}

This example is intentionally simplistic, but in reality, it is common to encounter hundreds of algorithmic variants with much longer kernel sequences for each variant. The manual process of making such discernments, even for this straightforward example, can be time-consuming and laborious. Therefore, given $\mathcal{M}_{gls}$ and the information regarding the kernel sequences in Table~\ref{tab3:gls-seq}, our objective is to automate the identification of the observations \textbf{r1}, \textbf{r2} and \textbf{r3} that we just manually discerned, and potentially uncover additional kernel sequence patterns that may have gone unnoticed during the manual analysis.

\begin{figure*}[h!]
	\centering
	\includegraphics[width=0.7\textwidth]{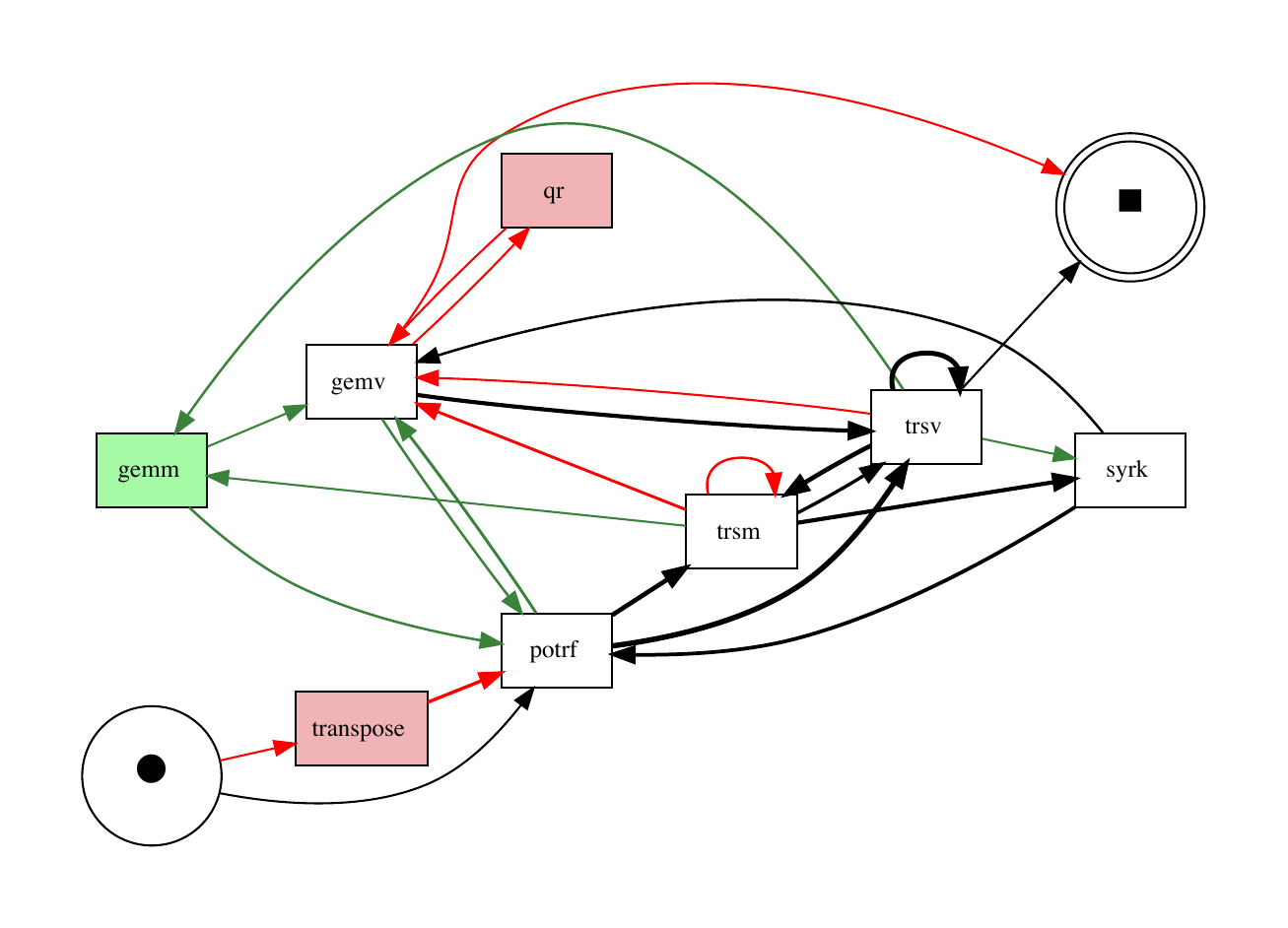}
	\caption{Directly Follows Graph: $\mathcal{G}reen$ = $\hat{\mathcal{R}}_0$, $\mathcal{R}ed$ = $\mathcal{\hat{R}}_1$.}
	\label{fig:dfg-f}
\end{figure*}

In Fig.~\ref{fig3:gls-eg}, we notice two well-separated groups (whether or not the data occurs in well-separated clusters can be automatically determined using a pre-trained binary classifier~\cite{kumari2017machine}), and we want to limit the number of ranks to the count of the well-separated groups, without unnecessarily creating additional ranks. Therefore, we apply Methodology~\ref{th:problem3} on  $\mathcal{M}_{gls}$ using the better-than relation $<_{eg}$, creating a partial ranking of the variants.
The resulting ranks ---$\mathcal{\hat{R}}_0$ and $\mathcal{\hat{R}}_1$--- are annotated in Fig.~\ref{fig3:gls-eg}. $\mathcal{\hat{R}}_0$ and $\mathcal{\hat{R}}_1$ are incidentally same as the manually identified Fast and Slow sets (shown in Equation~\ref{eq3:gls-gt}) respectively. 

\begin{figure*}[h!]
	\centering
	\includegraphics[width=0.7\textwidth]{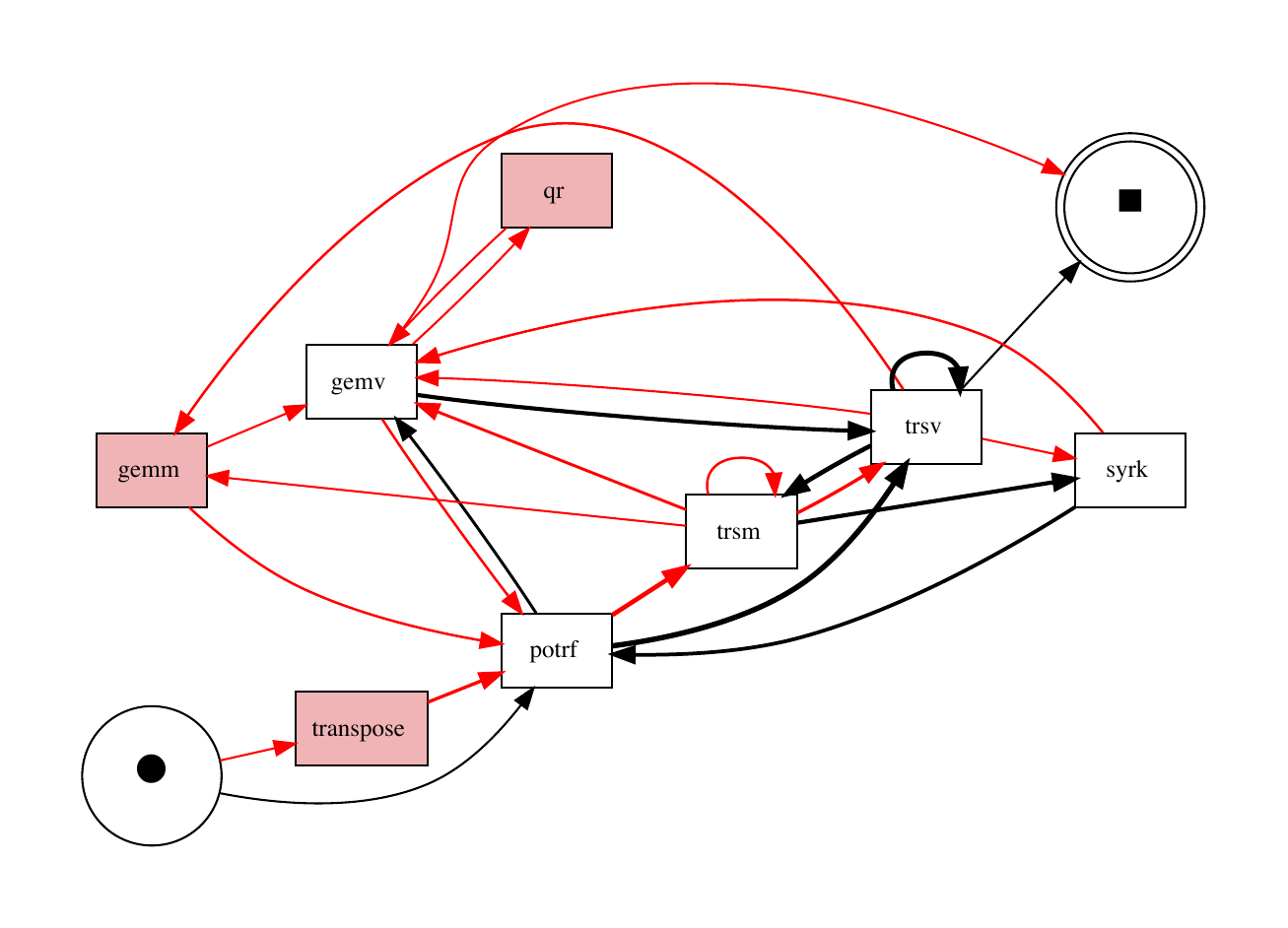}
	\caption{DFG (Top-1 median ranking):  $\mathcal{G}reen = \{ \mathbf{alg}_0\}$, $\mathcal{R}ed$ contains the remaining variants.}
	\label{fig3:dfg-m1}
\end{figure*}

In order to identify the similarities and differences among the variants in $\mathcal{\hat{R}}_0$ and $\mathcal{\hat{R}}_1$ in terms of kernel calls, we construct a Directly-Follows-Graph (DFG), where each node indicates a kernel, and a directed edge from one kernel to another, say from \textit{kernelA} to \textit{kernelB}, indicates that there exists a kernel sequence in which \textit{kernelA} directly precedes \textit{kernelB}. Given a split of the variants into two sets $\mathcal{G}reen$ and $\mathcal{R}ed$, we color the nodes and edges in the DFG such that green nodes and edges indicate the kernels and the directly-follows relations that occur \textit{exclusively} in the variants from $\mathcal{G}reen$. Similarly, red nodes and edges indicate the kernels and the directly-follows relations that occur exclusively in the variants from $\mathcal{R}ed$. All the other kernels and relations occur in the variants from both $\mathcal{G}reen$ and $\mathcal{R}ed$. The DFG with $\mathcal{G}reen = \hat{\mathcal{R}}_0$ and  $\mathcal{R}ed$ = $\mathcal{\hat{R}}_1$ is shown in Fig.~\ref{fig:dfg-f}. The kernel \textit{gemm} that occurs only in the variants from $\hat{\mathcal{R}}_0$ is colored green, and the kernels \textit{transpose} and \textit{qr} that occurs only in the variants from $\hat{\mathcal{R}}_1$ is colored red. Hence, we automatically identify the observations \textbf{r1}, \textbf{r2} and \textbf{r3}. We also discover additional patterns, such as the observation that whenever the kernels \textit{trsm} or \textit{trsv} directly precede \textit{gemv}, the corresponding variant is not one of the fast ones.

\begin{figure*}[h!]
	\centering
	\includegraphics[width=.7\textwidth]{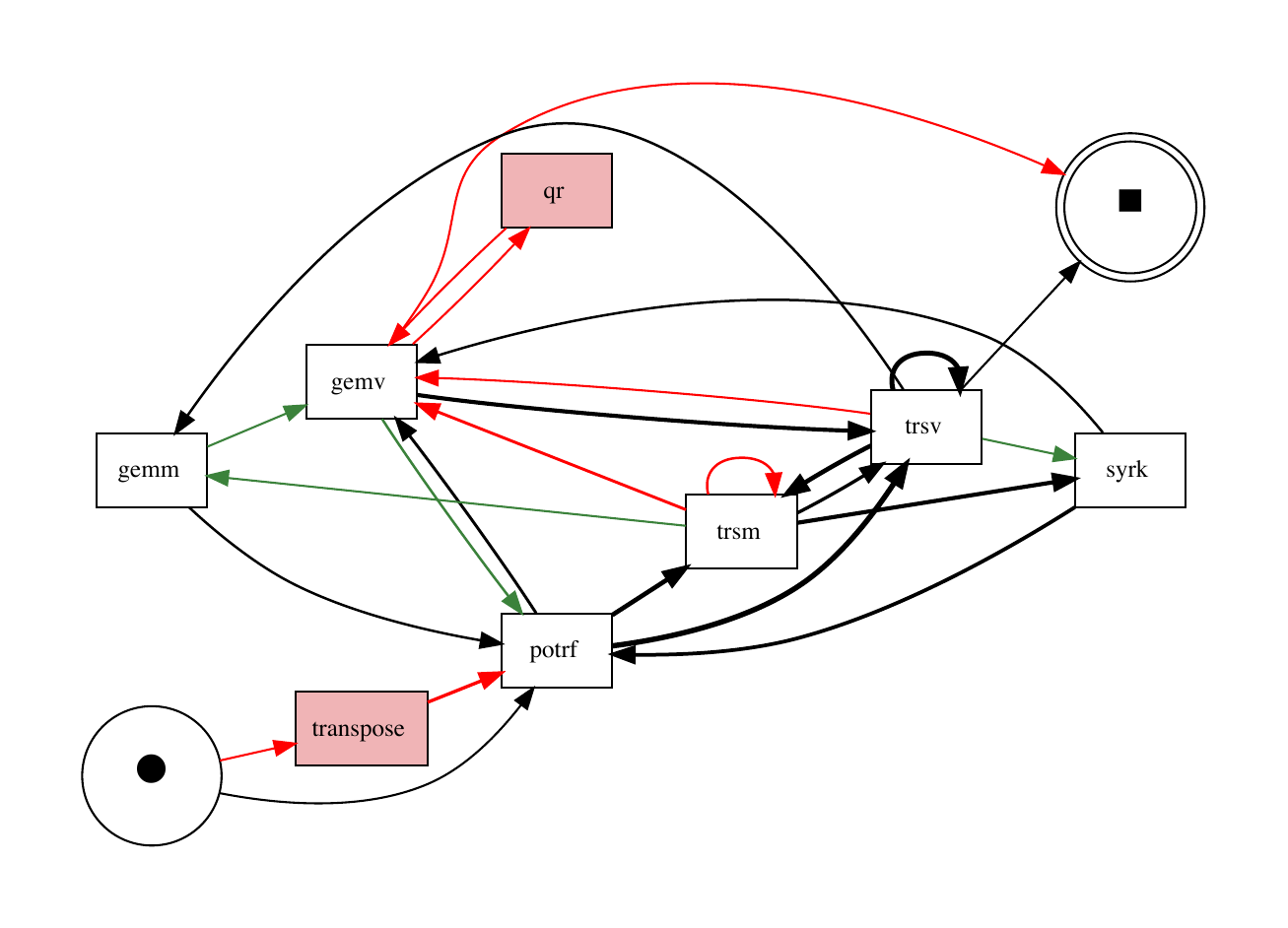}
	\caption{DFG (Top-4 median ranking):  $\mathcal{G}reen = \{ \mathbf{alg}_0, \mathbf{alg}_1, \mathbf{alg}_5, \mathbf{alg}_3\}$, $\mathcal{R}ed$ contains the remaining variants.}
	\label{fig3:dfg-m4}
\end{figure*}

In order to highlight the significance of using a ranking methodology that accounts for ties, we illustrate the limitations that are inherent when mining for the causes of performance difference based on rankings that determined by simply relying only on the median execution time of the variants. To this end, we consider the following:
\begin{itemize}
	\item \textbf{Top-1 median}: The variant having the lowest median execution time is placed in the set $\mathcal{G}reen$. The remaining variants are placed in the set $\mathcal{R}ed$. The resulting DFG is shown in the Fig.~\ref{fig3:dfg-m1}. As this kind of ranking completely ignores ties, the coloring of the DFG is not reasonable; the kernel \textit{gemm} is colored red because the only variant in the set $\mathcal{G}reen$, which is $\mathbf{alg}_0$, does not use this kernel.  However, there exists other variants --- $\mathbf{alg}_3$, $\mathbf{alg}_4$ and $\mathbf{alg}_5$ --- that use \textit{gemm} and the spread of their execution times significantly overlaps with that of $\mathbf{alg}_0$. Hence, according to the available data, it is not reasonable to highlight \textit{gemm} as a cause for a variant to be slow. 
	\item \textbf{Top-k median}: Top-k ranking is a common approach to distinguish between good and bad variants. Here, the variants with the top $k$ lowest median execution times are placed in the set $\mathcal{G}reen$ and the remaining variants are placed in the set $\mathcal{R}ed$. It is important to note that in this context, the identification of reasonable causes depends on the choice of k. For $k=5$, the variants are split into the two sets according to ones intuition (i.e., according to Equation~\ref{eq3:gls-gt}). However, for $k=4$, we get the DFG shown in Fig.~\ref{fig3:dfg-m4} in which the association \textbf{r1} is not identified. 
\end{itemize}

\subsection{Business Process Example}
\label{sec3:bpi}
Just as an algorithm can be viewed as a sequence of kernel calls, a business process can be viewed as a sequence of tasks or activities that need to be performed to achieve a specific business objective.
In large organizations, such processes are typically complex and hard to optimize, and enterprise tools are often used to monitor and improve such processes~\cite{van_der_aalst2016process}. We now show how partial ranking can be helpful in improving the insights provided by such tools.

Let us consider the Purchase-to-Pay (P2P)  business process, which involves the activities encountered while acquiring goods or services from external suppliers. For every purchase request, the sequence in which the activities are executed is captured  by sophisticated Enterprise Resource Management systems like SAP. A particular sequence of activities is referred to as a process variant, and in large organizations, there are typically hundreds of process variants for a business process like P2P (in the following, we show an example). 
\begin{figure*}[h!]
	\centering
	\includegraphics[width=0.8\textwidth]{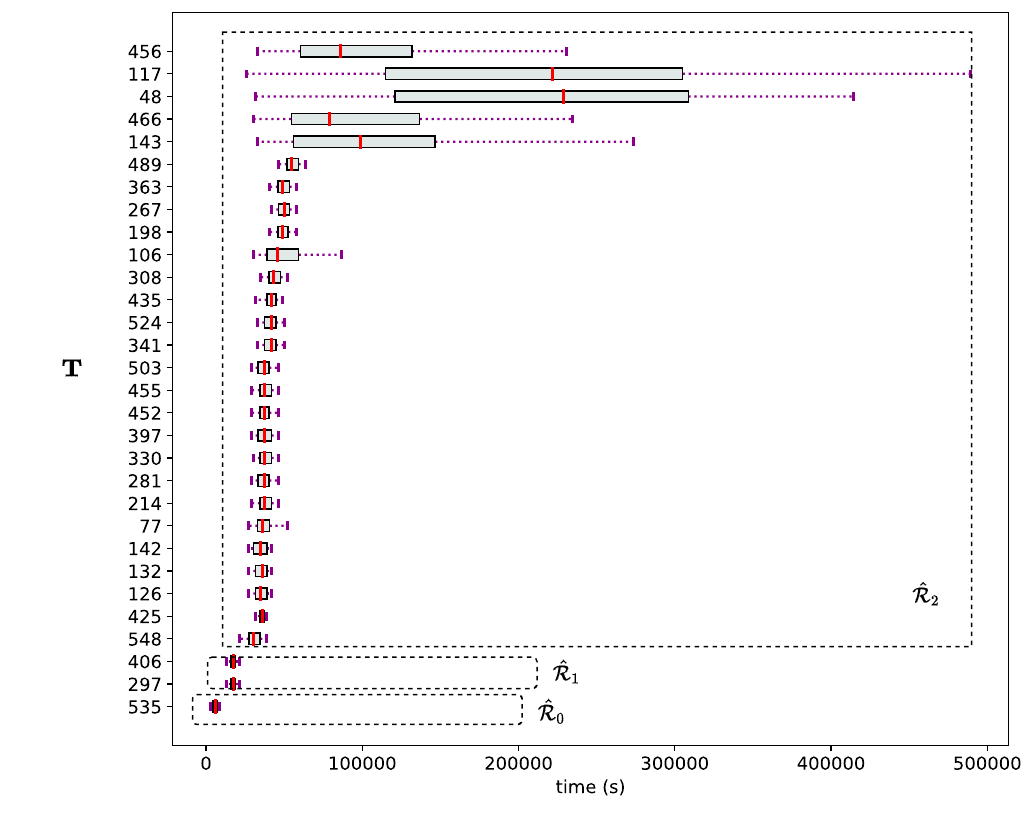}
	\caption{Sets of throughput times of the process variants ($\mathcal{M}_{bpi}$) and the partial ranking annotated based on Methodology~\ref{th:problem3} according to $<_{eg}$.}
	\label{fig3:bpi-eg}
\end{figure*}
\begin{figure*}[h!]
	\centering
	\includegraphics[width=\textwidth]{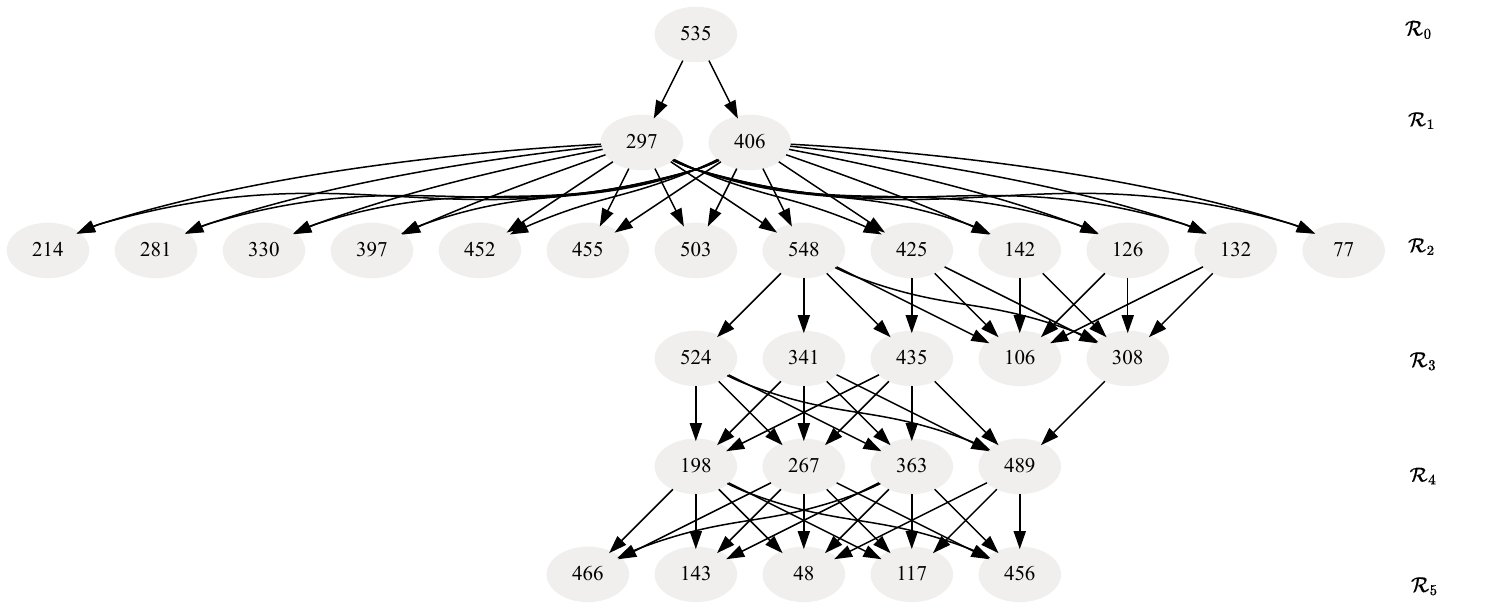}
	\caption{Methodology~\ref{th:problem2} on $\mathcal{M}_{bpi}$ according to $<_{eg}$. The graph $H$ is shown.}
	\label{fig3:bpi-eg-dfg}
\end{figure*}
For the purpose of illustration, we consider a P2P dataset from certain organization provided by a German data processing company - Celonis, 
during a hackathon event organized in collaboration with RWTH Aachen University in April 2022. 

\begin{figure*}[h!]
	\centering
	\includegraphics[width=\textwidth]{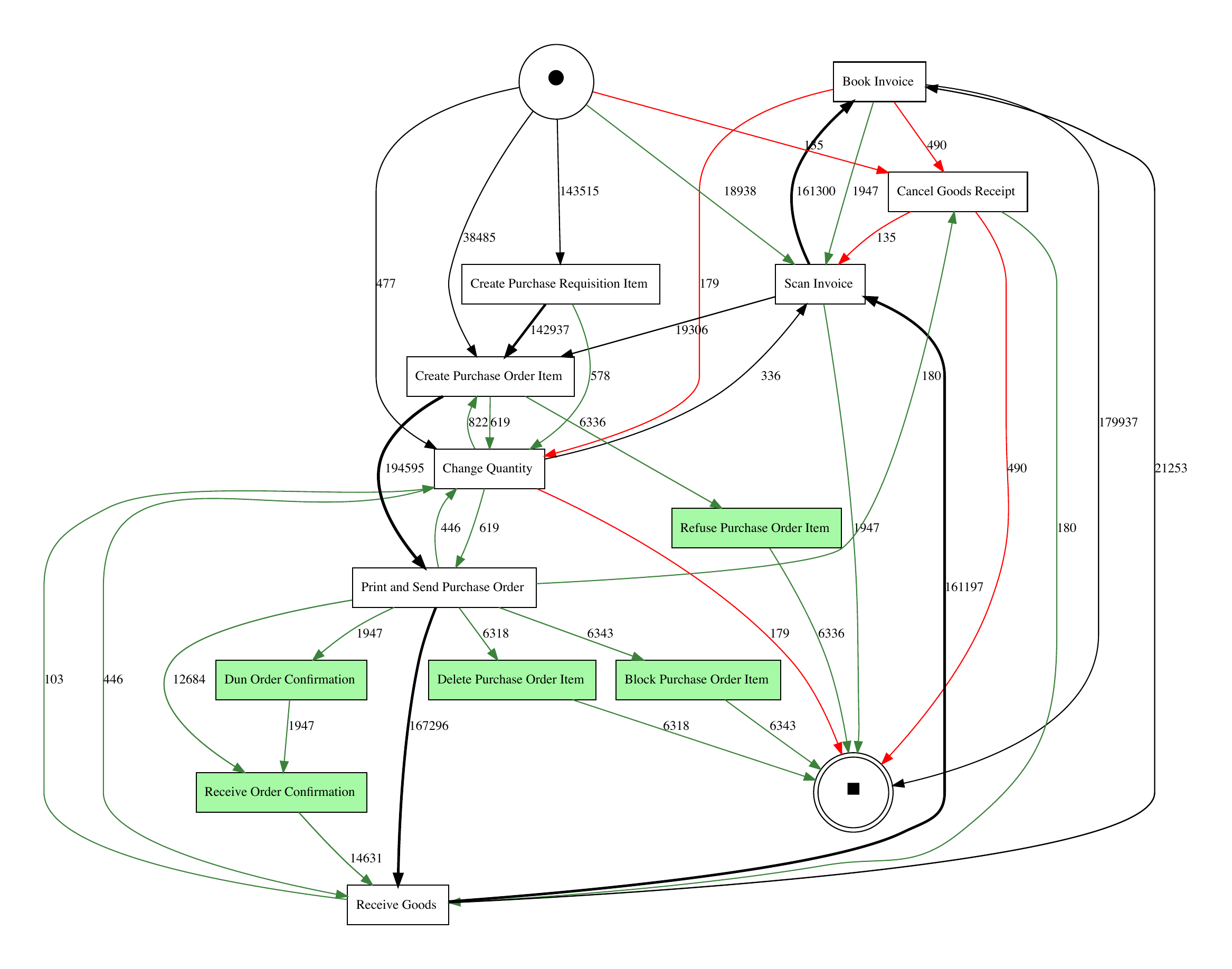}
	\caption{The DFG highlighting the root causes of performance differences between the process variants. The set $\mathcal{G}reen$ constitutes $\mathcal{R}_0$, $\mathcal{R}_1$ and $\mathcal{R}_2$, and the set $\mathcal{R}ed$ constitutes $\mathcal{R}_5$.}
	\label{fig3:bpi-eg-vc}
\end{figure*}

In the considered dataset,  the sets of throughput times of  the 30 most frequent process variants are shown in Fig.~\ref{fig3:bpi-eg} ($\mathcal{M}_{bpi}$). The partial ranking according to Methodology~\ref{th:problem3} with $<_{eg}$ is annotated in Fig.~\ref{fig3:bpi-eg}. However, as the spreads of the throughput times are largely overlapping with monotonically increasing differences, Methodology~\ref{th:problem3} and \ref{th:problem4} merges variants with significantly different performance into the same rank $\mathcal{\hat{R}}_2$. Moreover, the throughput times of the variants do not occur in well-separated groups. Therefore, instead of Methodology~\ref{th:problem3}, we employ Methodology~\ref{th:problem2} to calculate an alternate ranking that classifies the variants into six ranks $\mathcal{R}_0, \dots, \mathcal{R}_5$ as shown in Fig.~\ref{fig3:bpi-eg-dfg}, and perform the root cause analysis. To this end, we construct a DFG similar to the one created in Sec.~\ref{sec3:gls}, but this time with nodes representing  activities in the P2P process. The set $\mathcal{G}reen$ constitutes the process variants from the ranks $\mathcal{R}_0$, $\mathcal{R}_1$, $\mathcal{R}_2$ and the set $\mathcal{R}ed$ constitutes the process variants from the rank $\mathcal{R}_5$. Thus the green nodes and edges indicate the activities and relations that occur exclusively in the process variants of the set $\mathcal{G}reen$, while the red nodes and edges indicate the activities and relations that occur exclusively in the process variants of the set $\mathcal{R}ed$. The resulting DFG is shown in Fig.~\ref{fig3:bpi-eg-vc}. On the edges, we also indicate the number of times a particular directly-follows relation was observed. 

\begin{figure*}[h!]
	\centering
	\includegraphics[width=\textwidth]{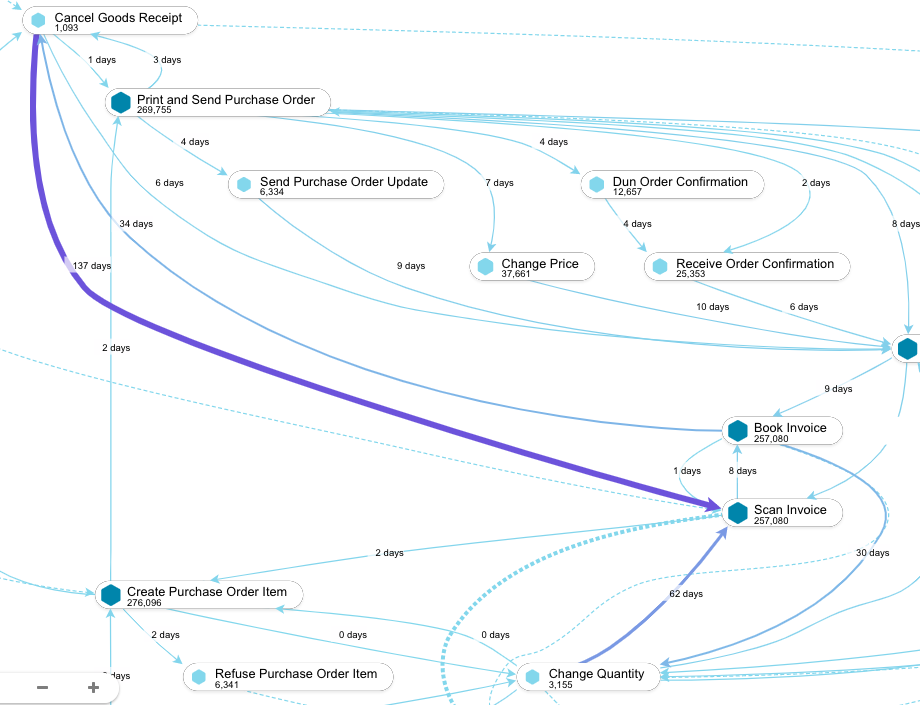}
	\caption{Screenshot of the DFG rendered by Celonis.}
	\label{fig3:bpi-eg-celonis}
\end{figure*}

We compare our constructed DFG with the DFG generated by the Celonis application, noting that the Celonis DFG does not apply partial ranking to classify and colors the nodes and edges based on the performance classes. A screenshot of a portion of the Celonis DFG is displayed in Fig.~\ref{fig3:bpi-eg-celonis}. In the Celonis DFG, the numbers displayed on the edges represent the median time between the start of two activities. Notably, there are significant time delays observed between the activities `Cancel Goods Receipt' to `Scan Invoice' (\textbf{e1}), `Change Quantity' to `Scan Invoice' (\textbf{e2}), and `Book Invoice' to `Change Quantity' (\textbf{e3}), indicating potential bottlenecks in the process flow. It is important to note that if a bottleneck occurs consistently across all process variants, it may not provide insights into the root cause of performance differences among the variants. However, when a bottleneck is present in some variants and not in others, it becomes a focal point for highlighting performance differences. While the Celonis DFG effectively captures these bottlenecks, it does not explicitly indicate whether these bottlenecks are the underlying causes of performance disparities among the variants. For example, in our DFG, the edge \textbf{e2} with a median time of 62 days is not highlighted in red as it appears in variants from both $\mathcal{G}reen$ and $\mathcal{R}ed$. Therefore, it is not inferred to directly contribute to the root cause of performance differences among the variants. However, the edge \textbf{e3}, with a median time of only 30 days, is marked in red in our DFG, suggesting it as an indicator of performance difference.

Thus, we remark that Celonis can benefit from incorporating partial ranking to provide additional perspectives within their DFG. This enhancement has the potential to offer valuable insights to their customers, aiding them in their decision-making processes.

\section{Conclusion}
\label{sec3:con}

We considered the problem of ranking sets of noisy measurement data while accounting for ties. As soon as ties are allowed, more than one reasonable ranking became possible because of the non-transitive nature of the ties. For given sets of measurements and a  better-than relation that defines how two sets of measurements should be compared, we defined partial ranking to identify a set of reasonable rankings. We formalized and developed three different methodologies for partial ranking. Methodology~\ref{th:problem2} computes the partial ranking consisting of an arbitrary number of ranks. Methodology~\ref{th:problem3} takes the partial ranking computed by Methodology~\ref{th:problem2} as input and aims to reduce the number of ranks. This methodology identifies an alternate partial ranking and does not necessarily compute the minimum possible number of ranks. 
Finally, we presented Methodology~\ref{th:problem4} which computes the partial ranking with minimum possible number of ranks. Methodology~\ref{th:problem3} can be seen as a trade-off between Methodology~\ref{th:problem2} and Methodology~\ref{th:problem4}, and particularly helpful when used as a starting point, especially when the number of measurements for the objects are not consistent.

We then demonstrated the application of partial ranking in discerning observations that could facilitate in identifying the causes of performance differences among the variants. We used the three methodologies interactively for the root cause analyses. An avenue for further extension of this research would be to develop a strategy to automatically identify the methodology that would be optimal for root cause analyses for a given dataset.

\bmhead{Acknowledgements}

Financial support from the Deutsche Forschungsgemeinschaft (German Research Foundation) through the grant IRTG 2379 is gratefully acknowledged. We also thank Celnois for granting access to the SAP P2P dataset through their cloud service, which enabled the execution of experiments in Sec.~\ref{sec3:bpi}. 

\bmhead{Availability of Data and Materials}

The experimental data and the results that support the findings of this study are available in Zenodo with the identifier \href{https://doi.org/10.5281/zenodo.12082779}{https://doi.org/10.5281/zenodo.12082779}. 

%
%
%
%


%

\bibliography{general,mypub,ranking}

\end{document}